\documentclass[11pt]{article}
\usepackage{colordvi}
\usepackage{epsfig}
\usepackage{axodraw}
\usepackage{epsfig}
\usepackage{graphicx}
\usepackage{rotate}
\usepackage{latexsym}
\usepackage{amssymb}
\usepackage{amsmath}
\usepackage{multirow}
\usepackage{dsfont}
\newcommand\pubnumber{TTP11-08}
\newcommand\pubdate{\today}

\def\csuma{Institut f\"ur Theoretische Teilchenphysik,
           Universit\"at Karlsruhe,\\
           76128 Karlsruhe, Germany}
\def\csumb{Department of Physics, University Of Alberta,\\
           Edmonton, AB T6G 2J1, Canada}
\def\csumc{Institute for Nuclear Research, Russian Academy of Sciences, \\
           117312 Moscow, Russia}
\def\csumd{Lehrstuhl f\"ur Theoretische Physik II, 
           Universit\"at W\"urzburg,\\
           97074 W\"urzburg, Germany}

\def\Title#1{\begin{center} {\Large\bf #1 } \end{center}}

\def\Author#1{\begin{center}{ \sc #1} \end{center}}
\def\Address#1{\begin{center}{ \it #1} \end{center}}

\newcommand\pubblock{\rightline{\begin{tabular}{l} \pubnumber\\
         \pubdate\\ \end{tabular}}}
\newenvironment{Abstract}{\begin{quotation}  }{\end{quotation}}

\def\Acknowledgments{\bigskip  \bigskip \begin{center}
          \large\bf Acknowledgments\end{center}}

\makeatletter
\def\section{\@startsection{section}{0}{\z@}{5.5ex plus .5ex minus
 1.5ex}{2.3ex plus .2ex}{\large\bf}}
\def\subsection{\@startsection{subsection}{1}{\z@}{3.5ex plus .5ex minus
 1.5ex}{1.3ex plus .2ex}{\normalsize\bf}}
\def\subsubsection{\@startsection{subsubsection}{2}{\z@}{-3.5ex plus
-1ex minus  -.2ex}{2.3ex plus .2ex}{\normalsize\sl}}

\renewcommand{\@makecaption}[2]{%
   \vskip 10pt
   \setbox\@tempboxa\hbox{\small #1: #2}
   \ifdim \wd\@tempboxa >\hsize     
       \small #1: #2\par          
     \else                        
       \hbox to\hsize{\hfil\box\@tempboxa\hfil}
   \fi}

 \def\citenum#1{{\def\@cite##1##2{##1}\cite{#1}}}
\def\citea#1{\@cite{#1}{}}
\newcount\@tempcntc
\def\@citex[#1]#2{\if@filesw\immediate\write\@auxout{\string\citation{#2}}\fi
  \@tempcnta\z@\@tempcntb\m@ne\def\@citea{}\@cite{\@for\@citeb:=#2\do
    {\@ifundefined
       {b@\@citeb}{\@citeo\@tempcntb\m@ne\@citea\def\@citea{,}{\bf }\@warning
       {Citation `\@citeb' on page \thepage \space undefined}}%
    {\setbox\z@\hbox{\global\@tempcntc0\csname b@\@citeb\endcsname\relax}%
     \ifnum\@tempcntc=\z@ \@citeo\@tempcntb\m@ne
       \@citea\def\@citea{,}\hbox{\csname b@\@citeb\endcsname}%
     \else
      \advance\@tempcntb\@ne
      \ifnum\@tempcntb=\@tempcntc
      \else\advance\@tempcntb\m@ne\@citeo
      \@tempcnta\@tempcntc\@tempcntb\@tempcntc\fi\fi}}\@citeo}{#1}}
\def\@citeo{\ifnum\@tempcnta>\@tempcntb\else\@citea\def\@citea{,}%
  \ifnum\@tempcnta=\@tempcntb\the\@tempcnta\else
  {\advance\@tempcnta\@ne\ifnum\@tempcnta=\@tempcntb \else\def\@citea{--}\fi
    \advance\@tempcnta\m@ne\the\@tempcnta\@citea\the\@tempcntb}\fi\fi}
\makeatother
\newcommand{\nl}{\nonumber\\}
\newcommand{\nn}{\nonumber}
\newcommand{\bq}{\begin{equation}}
\newcommand{\eq}{\end{equation}}
\newcommand{\bqa}{\arraycolsep 0.14em\begin{eqnarray}}
\newcommand{\eqa}{\end{eqnarray}}
\newcommand{\bqaa}{\begin{eqnarray*}}
\newcommand{\eqaa}{\end{eqnarray*}}
\newcommand{\ba}[1]{\begin{array}{#1}}
\newcommand{\ea}{\end{array}}
\newcommand{\bei}{\begin{itemize}}
\newcommand{\eei}{\end{itemize}}
\newcommand{\eqn}[1]{Eq.(\ref{#1})}

\newcommand{\fig}[1]{Fig.~\ref{#1}}

\newcommand{\sect}[1]{Section~\ref{#1}}

\newcommand{\be}{\begin{equation}}
\newcommand{\ee}{\end{equation}}
\newcommand{\bfm}[1]{\mbox{\boldmath$#1$}}

\newcommand{\al}{\alpha}

\newcommand{\lm}{\lambda}
\newcommand{\gm}{\gamma}

\newcommand{\dl}{\delta}

\newcommand{\dd}{\mbox{d}}

\newcommand{\gsim}{\;\rlap{\lower 3.5 pt \hbox{$\mathchar \sim$}} \raise 1pt \hbox {$>$}\;}
\newcommand{\lsim}{\;\rlap{\lower 3.5 pt \hbox{$\mathchar \sim$}} \raise 1pt \hbox {$<$}\;}
\newcommand{\sw}{s_{\!_W}}

\newcommand{\tw}{t_{\!_W}}
\newcommand{\shat}{\hat{s}}

\newcommand{\sigmahat}{\hat{\sigma}}
\newcommand{\pt}{p_{\rm T}}
\newcommand{\I}{\bfm{\mathds{1}}}

\newcommand{\ssSU}{\scriptscriptstyle{S\!U\!(\!2)}}
\newcommand{\ssY}{\scriptscriptstyle{Y}}
\newcommand{\ssQED}{{\scriptscriptstyle{Q\!E\!D}}}

\newcommand{\ssT}{{\scriptscriptstyle{T}}}
\newcommand{\ssL}{{\scriptscriptstyle{L}}}
\newcommand{\ssP}{{\scriptscriptstyle{P}}}
\newcommand{\wz}{w_{\!_Z}}
\newcommand{\wh}{w_{\!_H}}
\newcommand{\wt}{w_{\!_t}}
\newcommand{\betaz}{\beta_{\!_Z}}
\newcommand{\betah}{\beta_{\!_H}}

\newcommand{\Ltw}{L_{_{\!tW}}}
\newcommand{\Lz}{L_{\!_Z}}
\newcommand{\Lh}{L_{\!_H}}
\newcommand{\Lt}{L_{\!_t}}
\newcommand{\Slash}[1]{/\!\!\!{#1}}
\newcommand{\st}{s_\theta}
\newcommand{\ct}{c_\theta}

\begin{document}
\begin{titlepage}
\pubblock
\vfill
\Title{Next-to-Next-to-Leading Electroweak Logarithms \\[+.3cm]
for W-Pair Production at LHC}
\vfill
\Author{
J.H. K\"uhn $^a$,
F. Metzler $^a$,
A.A. Penin $^{a,b,c}$,
S. Uccirati $^d$
}

\Address{\csuma}
\Address{\csumb}
\Address{\csumc}
\Address{\csumd}

\vfill
\vfill
\begin{center}
{\bf Abstract}
\end{center}
\begin{Abstract}
\noindent
We derive the  high energy asymptotic of one- and two-loop corrections in the
next-to-next-to-leading logarithmic approximation  to the differential cross
section of \mbox{$W$}-pair production at the LHC.
For large invariant mass of the W-pair the (negative) one-loop terms can
reach more than $40\%$, which are partially compensated by the (positive)
two-loop terms of up to $10\%$.
\end{Abstract}
\vfill
\begin{center}
\\[5mm]
\end{center}
\end{titlepage}
\def\thefootnote{\arabic{footnote}}
\setcounter{footnote}{0}
\section{Introduction}
With the LHC now starting its operation, the experimental investigation
of scattering processes at the TeV scale is within reach.
Starting from these energies electroweak corrections are strongly enhanced by
Sudakov logarithms of the form $(\alpha_{ew}^n \ln^{2n} s/M_W^2)^n$ \cite{Fad}.
The full evaluation of electroweak one-loop corrections to fermion- or
$W$-pair production is by now a straightforward task.
Two-loop corrections, however,  can be obtained only in the high energy limit.
By employing the evolution equation approach the analysis of the dominant
logarithmically enhanced two-loop corrections for four-fermion processes has
been pushed successfully from next-to-leading logarithmic (NLL) approximation
\cite{KPS,Mel1,DMP} to  NNLL \cite{KMPS} and even  N${}^3$LL approximation
\cite{FKPS,JKPS}, which accounts for all the two-loop logarithmic terms (for
additional work on this topic see
e.g.~\cite{Kuhn:1999de,Ciafaloni:2000df,Denner:2000jv}).
Subsequent analysis  performed in the effective theory framework \cite{Man1}
employing the two-loop anomalous dimensions calculated in Refs.\cite{FKPS,JKPS}
have confirmed the formentioned result. 

In this paper we consider specifically pair production of $W$-bosons.
Previously, the electroweak corrections were studied mainly
in the context of the electron-positron annihilation.
The one-loop corrections have been evaluated  for the $W$-pair production
\cite{LemVel,Boh,FJZ,Bee} and the $W$-boson mediated $e^+e^-\to 4f$
processes ~\cite{Bee2,Jad,Den2,Den}.  For high energies the
two-loop logarithmically enhanced terms have been obtained
up to the NNLL  approximation \cite{Mel1,DMP,BRV,Kuhn:2007ca}.
The one-loop contribution amounts to typically -20\% for 1 TeV and -50\% for 3
TeV while  two-loop terms vary between 2 and 5\% for 1 TeV, for 3 TeV they may
even rise to 20\%. For the $W$-pair production at the LHC the  analysis of 
the one-loop electroweak logarithms to the NLL approximation is given in \cite{ADP,ADK}
with the  realistic cuts and the effect of gauge boson decay included. 
Beyond one loop the logarithmic corrections to the partonic  cross sections 
were considered in \cite{Man2}. In view of the extremely large partonic energies 
and with the LHC eventually operating at full luminosity (not to speak of the SLHC) 
invariant masses of the $W$-pair exceeding 1 TeV and approaching  3 TeV seem within reach. Therefore
the evaluation of the enhanced electroweak corrections  is
of particular interest. Here we present the explicit result for the one- and
two-loop  corrections to the partonic $q\bar q \to W^+W^-$ and
hadronic $p p \to  W^+W^- $ cross section in high energy limit in the NNLL
approximation. Note that the cross section of $W$-pair hadronic production is a subject of large
corrections due to the strong interaction of the  initial states. 
Currently  the analysis of  QCD  corrections is completed to the  NLO and NLL   
approximation (see \cite{BHO,DKS,Gra} and references therein). 
The size of the corrections depends strongly on a particular observable
and in many cases the available approximation provides a few percent accuracy. 
As we will see the two-loop  electroweak logarithms  become essential 
at this level of precision and have  to be included in the theoretical predictions.

Our paper is organized as follows: the partonic
processes in Born approximation are introduced in Section~\ref{sec2}.
In Section~\ref{sec3} the evolution equation approach is outlined  for the
simplified case of a pure $SU(2)$ spontaneously broken gauge theory.
The discussion closely follows  Ref.~\cite{Kuhn:2007ca}. However in the
present paper we derive the explicit result for the one-loop corrections to
scattering amplitudes given in  Appendix~A. The generalization to the
$SU(2)\otimes U(1)$ Standard Model is presented in Section~\ref{sec4}, which
contains a more detailed analysis of the  separation of infrared singularities
connected with virtual photon emission. The results  for the one- and
the two-loop corrections to the partonic cross section in NNLL approximation are
listed in the Appendix B. In Section~\ref{sec41} we present a numerical study of
these corrections for $\sqrt{\hat s}= 1$ TeV and 3 TeV
respectively. Based on these results, the corrections to the transverse and
longitudinal $W$-pair production in proton-proton collisions at 14~TeV are
presented in Section~\ref{sec42} together with the  discussion of the
anticipated statistical errors. Section~\ref{sec5} contains a brief summary and
conclusions. In Appendix~C we present the correction to the
two-loop NNLL result for the transverse  $W$-pair production in
electron-positron annihilation \cite{Kuhn:2007ca}.

\section{The partonic process}
\label{sec2}
\begin{figure}[ht]
$$
\scalebox{0.9}{
\begin{picture}(60,20)(0,-3)
 \ArrowLine(0,20)(30,20)        \Text(-5,17)[cb]{$d$}
 \ArrowLine(30,-20)(0,-20)      \Text(-5,-23)[cb]{$\bar{d}$}
 \ArrowLine(30,20)(30,-20)
 \Photon(60,20)(30,20){2}{6}    \Text(73,17)[cb]{$W^-$}
 \Photon(30,-20)(60,-20){2}{6}  \Text(73,-23)[cb]{$W^+$}
\end{picture}
}
\qquad\qquad\qquad
\scalebox{0.9}{
\begin{picture}(60,20)(0,-3)
 \ArrowLine(0,20)(30,20)        \Text(-5,17)[cb]{$u$}
 \ArrowLine(30,-20)(0,-20)      \Text(-5,-23)[cb]{$\bar{u}$}
 \ArrowLine(30,20)(30,-20)
 \Photon(60,20)(30,-20){2}{7}    \Text(73,17)[cb]{$W^-$}
 \Photon(30,20)(60,-20){2}{7}  \Text(73,-23)[cb]{$W^+$}
\end{picture}
}
\qquad\qquad\qquad
\scalebox{0.9}{
\begin{picture}(60,20)(0,-3)
 \ArrowLine(0,20)(15,0)        \Text(-5,17)[cb]{$q$}
 \ArrowLine(15,0)(0,-20)       \Text(-5,-23)[cb]{$\bar{q}$}
 \Photon(15,0)(45,0){2}{7}     \Text(30,5)[cb]{$\gamma,Z$}
 \Photon(60,20)(45,0){2}{5}    \Text(73,17)[cb]{$W^-$}
 \Photon(45,0)(60,-20){2}{5}   \Text(73,-23)[cb]{$W^+$}
\end{picture}
}
$$
\\[-.5cm]
\caption{Tree level diagrams contributing to the partonic process}
\label{fig:born}
\end{figure}
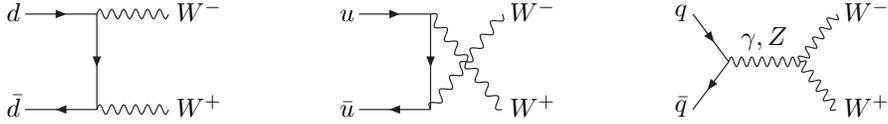

The partonic processes relevant for the $W$-pair production at hadron
colliders are gluon fusion and quark-antiquark annihilation.
The gluon contribution to the total cross section is about $5\%$ 
\cite{Binoth:2006mf} and we focus on the process $q\,\bar{q}\to W^+\,W^-$. 
In the leading order it is described by  the diagrams in \fig{fig:born}.
\vspace{0.5cm}

\begin{figure}[ht]
$$
\scalebox{0.9}{
\begin{picture}(60,20)(0,-3)
 \ArrowLine(0,20)(30,20)        \Text(-5,17)[cb]{$q$}
 \ArrowLine(30,-20)(0,-20)      \Text(-5,-23)[cb]{$\bar{q}$}
 \ArrowLine(30,20)(30,-20)
 \Photon(60,20)(30,20){2}{6}    \Text(73,17)[cb]{$W^-$}
 \Photon(30,-20)(60,-20){2}{6}  \Text(73,-23)[cb]{$W^+$}
\end{picture}
}
\qquad\qquad
\scalebox{0.9}{
\begin{picture}(60,20)(0,-3)
 \ArrowLine(0,20)(15,0)        \Text(-5,17)[cb]{$q$}
 \ArrowLine(15,0)(0,-20)       \Text(-5,-23)[cb]{$\bar{q}$}
 \Photon(15,0)(45,0){2}{7}     \Text(30,5)[cb]{$\gamma,Z$}
 \Photon(60,20)(45,0){2}{5}    \Text(73,17)[cb]{$W^-$}
 \Photon(45,0)(60,-20){2}{5}   \Text(73,-23)[cb]{$W^+$}
\end{picture}
}
\qquad\qquad
\scalebox{0.9}{
\begin{picture}(30,20)(0,-3)
\LongArrow(0,0)(30,0)
\Text(15,-15)[cb]{\small high energy}
\Text(15,-22)[cb]{\small limit}
\end{picture}
}
\qquad\quad
\scalebox{0.9}{
\begin{picture}(60,20)(0,-3)
 \ArrowLine(0,20)(30,20)        \Text(-5,17)[cb]{$q$}
 \ArrowLine(30,-20)(0,-20)      \Text(-5,-23)[cb]{$\bar{q}$}
 \ArrowLine(30,20)(30,-20)
 \Photon(60,20)(30,20){2}{6}    \Text(73,17)[cb]{$W_{_T}^-$}
 \Photon(30,-20)(60,-20){2}{6}  \Text(73,-23)[cb]{$W_{_T}^+$}
\end{picture}
}
\qquad\qquad
\scalebox{0.9}{
\begin{picture}(60,20)(0,-3)
 \ArrowLine(0,20)(15,0)        \Text(-5,17)[cb]{$q$}
 \ArrowLine(15,0)(0,-20)       \Text(-5,-23)[cb]{$\bar{q}$}
 \Photon(15,0)(45,0){2}{7}     \Text(30,5)[cb]{$\gamma,Z$}
 \DashLine(60,20)(45,0){3}     \Text(73,17)[cb]{$\phi^-$}
 \DashLine(45,0)(60,-20){3}    \Text(73,-23)[cb]{$\phi^+$}
\end{picture}
}
$$
\\[-.5cm]
\caption{Goldstone equivalence theorem at Born level}
\label{fig:goldstone}
\end{figure}
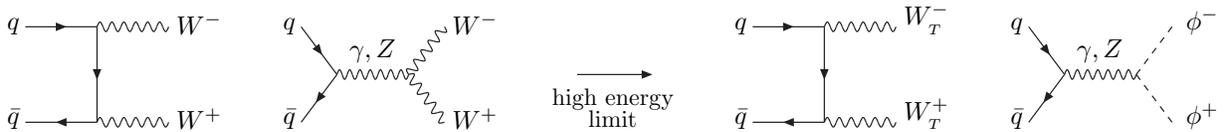

\noindent
The kinematics at partonic level is defined by:
\bqa
q(p_1,\lambda_+) + \bar{q}(p_2,\lambda_-) &\to&
W^+(k_+,\kappa_+) + W^-(k_-,\kappa_-),
\eqa
where $\lambda_\pm$ and $\kappa_\pm$ are the helicities of the incoming
and outgoing particles respectively.
For on-shell $W$-bosons, the matrix element can be then expressed as function
of the Mandelstam variables:
\bq
\hat{s}= (p_1+p_2)^2, \qquad
\hat{t}= (p_1-k_-)^2, \qquad
\hat{u}= (p_1-k_+)^2,
\eq
They are related to the scattering angle $\theta$ through the relations:
\bq
\hat{t}= M_W^2 - \frac{\hat{s}}{2}\Big( 1 - \beta\,\cos\theta \Big)\,,
\qquad\
\hat{u}= M_W^2 - \frac{\hat{s}}{2}\Big( 1 + \beta\,\cos\theta \Big)\,,
\qquad\quad
\beta^2= 1 - 4\,\frac{M_W^2}{\hat{s}}\,.
\eq
In the high energy limit only final states where the  $W$-bosons have the
same polarization are not suppressed by a factor $M_W^2/s$ or higher.
In addition, the case where both $W$'s are longitudinal can be reduced
by means of the Goldstone equivalence theorem to the production of a pair
of charged Goldstone bosons as shown in \fig{fig:goldstone}.
\vspace{-.5cm}

\section{Massive gauge boson production in $SU(2)$ model}
\label{sec3}
Let us, in a first step, neglect the hypercharge and consider a simplified
model with spontaneously broken gauge group $SU(2)$. The model retains
the main features of the massive gauge boson sector of the Standard Model.
In this case the result can be presented in a simple analytical form and
constitutes the basis for the further extension to the full electroweak theory.
We study the process of gauge boson pair production in  fermion-antifermion
annihilation at high energy and fixed angle with all kinematical
invariants of the same order and far larger than the gauge boson mass $M$,
$|s|\sim |t| \sim |u| \gg M^2$. In this limit the asymptotic energy
dependence of the amplitudes is dominated by {\em Sudakov}
logarithms  \cite{Sud,Jac} and  governed by the evolution
equations \cite{Mue,Col,Sen1}.
The method of the evolution equations in the context of the
electroweak corrections is described in detail for fermion pair
production in Ref.~\cite{JKPS} and for $W$-pair production in
Ref.~\cite{Kuhn:2007ca}.

\begin{figure}[t]
$$
{\cal F}_\psi
\;
=
\;
\raisebox{0.1cm}{\scalebox{0.9}{
\begin{picture}(80,30)(0,0)
 \ArrowLine(0,20)(30,10)
 \ArrowLine(30,-10)(0,-20)
 \GCirc(40,0){15}{0.8}
 \Photon(55,0)(80,0){2}{6}
 \LongArrow(78,7)(68,7)        \Text(85,4)[cb]{$p$}
\end{picture}
}}
\qquad\qquad\quad
{\cal F}_\phi
\;
=
\;
\raisebox{0.1cm}{\scalebox{0.9}{
\begin{picture}(80,30)(0,0)
 \DashLine(0,20)(30,10){3}
 \DashLine(0,-20)(30,-10){3}
 \GCirc(40,0){15}{0.8}
 \Photon(55,0)(80,0){2}{6}
 \LongArrow(78,7)(68,7)        \Text(85,4)[cb]{$p$}
\end{picture}
}}
\qquad\qquad\quad
{\cal F}_A
\;
=
\;
\raisebox{0.1cm}{\scalebox{0.9}{
\begin{picture}(80,30)(0,0)
 \Photon(0,20)(30,10){2}{6}
 \Photon(0,-20)(30,-10){2}{6}
 \GCirc(40,0){15}{0.8}
 \Line(55,0)(80,0)
 \LongArrow(78,7)(68,7)        \Text(85,4)[cb]{$p$}
\end{picture}
}}
$$
\\[-.8cm]
\caption{Fermion/scalar
scattering in an external singlet vector
field and scattering of a gauge boson in an external scalar field.
The momentum of external field satisfies $p^2=s=-Q^2$.
}
\label{Z}
\end{figure}
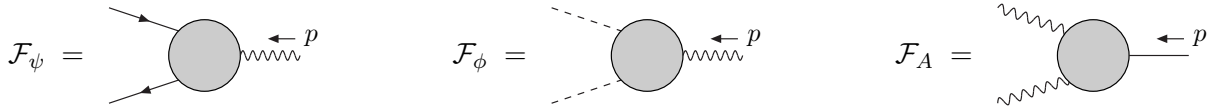
Following Ref.~\cite{Kuhn:2007ca} we introduce the functions ${\cal
Z}_{\psi,\phi,A}$ which describe the asymptotic dependence on the large momentum
transfer $Q$ of the  scattering amplitude of the spinor ($\psi$) or  scalar
($\phi$) field in an external singlet vector field and of the vector boson
($A$) in an external singlet scalar field, {\it i.e.} of the respective form
factors in the  Euclidean region (see \fig{Z}). In  leading order in $M^2/Q^2$
these functions are known to satisfy the following linear evolution equation
\cite{Mue,Col,Sen1}
\begin{eqnarray}
{\partial\over\partial\ln{Q^2}}{\cal Z}_i&=&
\left[\int_{M^2}^{Q^2}{\dd x\over x}\gm_i(\al(x))+\zeta_i(\al(Q^2))
+\xi_i(\al(M^2)) \right] {\cal Z}_i \,,
\label{evoleqz}
\end{eqnarray}
with the solution
\begin{eqnarray}
{\cal Z}_i&=&\exp \left\{\int_{M^2}^{Q^2}{\dd x\over x}
\left[\int_{M^2}^{x}{\dd x'\over x'}\gm_i(\al(x'))+\zeta_i(\al(x))
+\xi_i(\al(M^2))\right]\right\}
\,,
\label{evolsolz}
\end{eqnarray}
which satisfies the initial condition ${\cal Z}_i\big|_{Q^2=M^2}=1$.  Here the
perturbative functions $\gm_i(\al)$ {\it etc.} are given by the series
in the coupling constant $\alpha(\mu^2)$, {\it e.g.}
$\gm_i(\al)=\sum_{n=1}^\infty (\al/4\pi)^n\gm_i^{(n)} $.
Then the amplitude of the transverse (longitudinal) gauge boson production
$\bfm{\cal A}_T$ ($\bfm{\cal A}_L$)  can be decomposed as follows
\begin{equation}
\bfm{\cal
  A}_{T,L}=\alpha(\mu_{T,L}){\cal Z}_\psi{\cal Z}_{A,\phi}\bfm{\tilde{\cal A}}_{T,L}\,.
\label{ampred}
\end{equation}
where $\bfm{\tilde{\cal A}}_{T,L}$ is the reduced amplitude and we factor out
the Born coupling constant $\alpha(\mu_{T,L})$. The scale dependence of this
factor is cancelled by the higher order renormalization group logarithms
replacing $\mu_{T,L}$ by a physical scale of the process. In the case of the
longitudinal $W$-pair the proper scale is  $\mu_L=\sqrt{s}$ because it describes
the interaction of far off-shell intermediate gauge boson with virtuality of
order $\sqrt{s}$. For the transverse $W$-pair production it is  $\mu_T=M_W$ 
corresponding to the coupling of the on-shell $W$-bosons. Note that in an
alternative approach based on the soft-collinear effective theory \cite{Man2} 
the normalization scale of the Born coupling constant for the transverse gauge
bosons is set to  $\sqrt{s}$. This is compensated by an additional  $\beta_0$ 
contribution to the anomalous dimension $\zeta_A^{(1)}$ which effectively shift
the normalization   of the Born coupling constant to $M_W$, in agreement with
our result.

Due to the factorization property of the Sudakov logarithms associated
with the {\it collinear} divergences of the massless theory
\cite{FreTay} the reduced amplitude satisfies the simple renormalization
group like equation \cite{Sen2,Ste,BotSte}
\begin{equation}
{\partial \over \partial \ln{Q^2}}\bfm{\tilde{\cal A}}_{T,L}=
\bfm{\chi}_{T,L}(\al(Q^2))\bfm{\tilde{\cal A}}_{T,L}\,,
\label{evoleqa}
\end{equation}
where $Q^2=-s$ and  $\bfm{\chi}_{T,L}$ is the soft anomalous dimension matrix
acting in the space of the isospin amplitudes.  The solution of the
above equation is given by the path-ordered exponent
\begin{equation}
\bfm{\tilde{\cal A}}_{T,L}=
{\rm P}\!\exp{\left[\int_{M^2}^{Q^2}
{\dd x\over x}\bfm{\chi}_{T,L}(\al(x))\right]}{\bfm{\cal
A}_0}_{\,T,L}(\al(M^2))\,,
\label{evolsola}
\end{equation}
where ${\bfm{\cal A}_0}_{\,T,L}$ determines the initial conditions for the
evolution equation at $Q=M$. By calculating the functions entering the
evolution equations order by order in $\al$ one gets the logarithmic
approximations for the amplitude. By expanding the exponents
one gets  the one- and two-loop corrections in the following form
\bqa
\bfm{\cal A}_{\ssT\!,\ssL}^{(1)}
&=&
\bigg[
  {1 \over 2}\gamma^{(1)} L^2
+ \Big( \zeta^{(1)}\! + \!\xi^{(1)}\! + \!\bfm{\chi}_{\ssT\!,\ssL}^{(1)} \Big)L
  \bigg] \bfm{\cal A}_{\ssT\!,\ssL}^{(0)}
+ \bfm{\cal A}_{0,\ssT\!,\ssL}^{(1)},
\qquad\quad
f^{(1)}\!= f_\psi^{(1)}\! + \!f_{A,\phi}^{(1)}\,,
\qquad
f\!= \gamma,\zeta,\xi,
\nl
\bfm{\cal A}_{\ssT\!,\ssL}^{(2)}
&=&
  \Bigg\{
    {1 \over 8}\,\big[ \gamma^{(1)} \big]^2 L^4\,
  + \,\frac{1}{2}\gamma^{\!(1)}
    \bigg[
    \zeta^{(1)} + \xi^{(1)} + \bfm{\chi}_{\ssT\!,\ssL}^{(1)} - \frac{1}{3}\beta_0
    \bigg] L^3\,
  + \,{1 \over 2}
    \bigg[
      \gamma^{\!(2)}\!
    + \!\Big( \zeta^{(1)}\! + \xi^{(1)}\! + \bfm{\chi}_{\ssT\!,\ssL}^{(1)} \Big)^2
\nl
&&
    - \beta_0\Big(\zeta^{(1)}\! + \!\bfm{\chi}_{\ssT\!,\ssL}^{(1)} \Big)
    \bigg] L^2
  \Bigg\}\,\bfm{\cal A}_{\ssT\!,\ssL}^{(0)}\,
+ \,{1 \over 2}\gamma^{\!(1)} L^2 \bfm{\cal A}_{0,\ssT\!,\ssL}^{(1)}\,
+ \,{\cal O}(L),
\qquad\qquad
\gamma^{(2)}\!= \gamma_\psi^{(2)}\! + \!\gamma_{A,\phi}^{(2)}\,,
\label{expsola}
\eqa
where $L=\ln(Q^2/M^2)$ and $\beta_0$ is the one-loop beta function.
The anomalous dimensions $\gm(\alpha)$, $\zeta(\alpha)$ and
$\bfm{\chi}(\al)$ are mass-independent and can be associated with
the infrared divergences of the massless (unbroken) theory.
At the same time the functions $\xi_i(\al)$ and ${\cal A}_{0,\ssT,\ssL}(\al)$
do depend on the infrared structure of the model and require the calculation
in the spontaneously broken phase.
All the perturbative coefficients in Eqs.~(\ref{expsola}) except
$\bfm{{\cal A}}_{0,\ssT,\ssL}^{(1)}$   are known \cite{Kuhn:2007ca}.
In Ref.~\cite{Kuhn:2007ca} the result for the one-loop
correction to the cross section \cite{Den}
has been used to obtain the two-loop NNLL terms.
We complete this part of the calculation and present the explicit result for
the one-loop corrections to the amplitude in Appendix A.
Our result for the cross section agrees with Ref.~\cite{Den}.

The large Yukawa coupling of the third generation quarks to the
scalar (Higgs and Goldstone) bosons results in specific logarithmic corrections
proportional to  $m_t^2/M_W^2$. This kind of Sudakov logarithms were
studied in Ref.~\cite{Col} and have  universal structure for any renormalizable
non-gauge theory. The factorization  in this case is much simpler than in gauge
theories and  the logarithmic corrections are completely determined by the
ultraviolet field renormalization of the external on-shell lines.
Since the Yukawa coupling of the initial light quark states is suppressed the
Yukawa enhanced  Sudakov logarithms for hadronic production of $W$-pair are
similar to those for $W$-pair production in electron-positron annihilation
\cite{Kuhn:2007ca}. Thus the Yukawa enhanced  corrections can be taken into
account through the modification of the evolution equations for the
corresponding $\cal Z_\phi$-function.
The main complication  is that the Yukawa interaction mixes the evolution of
the quark and scalar boson form factors and in general does not commute with
the $SU(2)$ coupling.
Thus the evolution equation has a complicated matrix form:
\begin{equation}
{\partial\over\partial\ln{Q^2}}\bfm{\cal Z}=
\Bigg[\int_{M^2}^{Q^2}{\dd x\over
x}\bfm{\gm}(\alpha(x))+\bfm{\zeta}(\alpha(Q^2),\alpha_{Y\!uk}(Q^2))
+\bfm{\xi}(\alpha(M^2)) \Bigg] \bfm{\cal Z} \,,
\label{evoleqzvec}
\end{equation}
with the solution
\begin{equation}
\bfm{\cal Z}={\rm P}\!\exp \left\{\int_{M^2}^{Q^2}{\dd x\over x}
\left[\int_{M^2}^{x}{\dd x'\over
x'}\bfm{\gm}(\al(x'))+\bfm{\zeta}(\alpha(x),\al_{Y\!uk}(x))
+\bfm{\xi}(\al(M^2))\right]\right\}\bfm{\cal Z}_0 \,,
\label{evolsolzvec}
\end{equation}
where  $\bfm\gamma^{(1)}=(-3/2)\bfm{1}$, $\bfm{\xi}=0$,
$\al_{Y\!uk}= M_t^2\big/(2M_W^2)\,\alpha$,  and we introduce the five-component
vector
\begin{equation}
\bfm{\cal Z}=({\cal Z}_\phi,~{\cal
Z}_\chi,~{\cal Z}_{b-},~{\cal Z}_{t-},~{\cal Z}_{t+})\,.
\end{equation}
The subscript $+$ $(-)$ stand for the right (left) quark  fields and
${\cal Z}_\chi$ corresponds to the transition of the Higgs boson
into the neutral Goldstone boson in the external singlet vector
field. The one-loop anomalous dimension matrix reads
\cite{Kuhn:2007ca}
\begin{equation}
\bfm{\zeta}^{(1)}={1\over 4}\begin{pmatrix} 12 & 0 & 0 & 0 & 0\cr
           0 & 12 & 0 & 0 &  0 \cr
           0& 0 & {9} & 0 & 0\cr
           0 & 0& 0 & {9} &  0 \cr
           0& 0 & 0&0 &{0} \cr
           \end{pmatrix}\,+{\rho\over 2}
\begin{pmatrix}  0 & 0 & {6} & 0 &
-{6}\cr
           0 & 0 & 0 & {6} & -{6}\cr
           {1}& 0 & 0 & 0 & -{1}\cr
           0 & {1}& 0 & 0& -{1} \cr
           -{1} & -{1}  & -{1} &-{1}  &0 \cr
           \end{pmatrix}\,,
\label{1lzetayuk}
\end{equation}
where the first term represents the  pure $SU_L(2)$ contribution, the second
term represents the Yukawa contribution and we introduce the ratio
$\rho = \alpha_{Y\!uk}/\alpha = \frac{M_t^2}{2M_W^2} \sim 1$.
The proper initial condition for the evolution equation which
corresponds to the $SU_L(2)$ Born amplitudes of the third generation quark and scalar boson
production in light quark-antiquark annihilation is given by the vector
$\bfm{\cal Z}_0=2(T^3_\phi,~T^3_\chi,~T^3_{b-},~T^3_{t-},~T^3_{t+})=(1,-1,-1,1,0)$ where 
$T^3_i$ stands for the particle isospin and the overall factor of 2 is introduced for convenience. 
Since the Yukawa enhanced logarithmic corrections can be attributed to the 
external on-shell field renormalization we expect a diagonal form of the corrections. 
This is indeed the case due to  a nontrivial matrix relation 
\begin{equation}
\left(\bfm{\zeta}^{(1)}_{Y\!uk}\right)^{2n}\cdot\bfm{\cal
Z}_0=\left({3\rho^2\over 2}\right)^n\bfm{\cal
Z}_0,\qquad  \left(\bfm{\zeta}^{(1)}_{Y\!uk}\right)^{2n+1}\cdot\bfm{\cal
Z}_0=\left({3\rho^2\over 2}\right)^n\left(-3\rho,3\rho,\rho/2,-\rho/2,0\right)\,,
\end{equation}
where the  vector on the right hand side of the second equation
represents the one loop  correction $\bfm{\zeta}^{(1)}_{Y\!uk}\cdot\bfm{\cal
Z}_0$.  By factorizing the components of $\bfm{\cal Z}_0$ we can rewrite it as follows
\begin{equation}
\bfm{\zeta}^{(1)}_{Y\!uk}\cdot\bfm{\cal
Z}_0=2\rho(-3 T^3_\phi,-3 T^3_\chi,-{1\over 2}T^3_{b-},-{1\over 2}T^3_{t-},-T^3_{t+})\,.
\label{1lyuk}
\end{equation}
The coefficients of $T^3_i$ in the above expression  depend only on the  field renormalization
of the particle $i$  as it has been explicitly shown in Ref.~\cite{Mel2}.
For example, for the hypercharge mediated Born amplitudes of the same process we have  
different initial conditions $\bfm{\cal Z}_0=(Y_\phi,~Y_\chi,~Y_{b-},~Y_{t-},~Y_{t+})=(1,1,1/3,1/3,4/3)$ 
where  $Y_i$ stands for the particle hypercharge. However the one loop coefficients 
are the same as in Eq.~(\ref{1lyuk})
\begin{equation}
\bfm{\zeta}^{(1)}_{Y\!uk}\cdot\bfm{\cal Z}_0=\rho(-3Y_\phi,-3Y_\chi,-{1\over 2}Y_{b-},-{1\over 2}~Y_{t-},-Y_{t+})\,.
\end{equation}
The different form of the odd and even order corrections is dictated by the off-diagonal 
character  of the  matrix of  field renormalization by Yukawa interaction. 

By expanding the solution for  the component
${\cal Z}_\phi$ we obtain the Yukawa enhanced logarithmic corrections
to the amplitude of the longitudinal $W$-pair production. Let us introduce the
following notation
\begin{equation}
\langle\bfm{\zeta}\rangle_{Y\!uk}=\left[\bfm{\zeta}\cdot\bfm{\cal
Z}_0\right]_\phi\,,
\end{equation}
where only the terms proportional to the second or fourth power of
the top quark mass are kept on the right hand side.
Then the Yukawa contribution to the amplitude~(\ref{expsola}) takes the
following form
\bqa
\bfm{\cal A}^{(1)}\big|_{Y\!uk} &=&
  \langle\bfm{\zeta}^{(1)}\rangle_{Y\!uk} L\, \bfm{\cal A}^{(0)}
+ \bfm{{\cal A}}_{0}^{(1)}\big|_{Y\!uk}
\nl
\bfm{\cal A}^{(2)}|_{Y\!uk} &=&
  \Bigg\{
    \frac{1}{2}\gamma^{\!(1)}\langle\bfm{\zeta}^{(1)}\rangle_{Y\!uk} L^3
  + \bigg[
      \bigg(
      \zeta^{(1)} + \xi^{(1)} + \bfm{\chi}^{(1)} - {1\over2}\beta_0^{Y\!uk}
      \bigg)
      \langle\bfm{\zeta}^{(1)}\rangle_{Y\!uk},
\nl
&&
    + {1\over2}\langle\big(\bfm{\zeta}^{(1)}\big)^2\rangle_{Y\!uk}
    \bigg] L^2
  \Bigg\} \bfm{\cal A}^{(0)}
+ {1 \over 2}\gamma^{\!(1)} L^2
  \bfm{{\cal A}}_{0}^{(1)}\big|_{Y\!uk}
+ {\cal O}(L),
\label{yukawa}
\eqa
where $\beta_0^{Y\!uk}=9/4-3\rho/2$ is the one-loop
beta-function of the Yukawa coupling constant and $\bfm{{\cal
A}}_{0}^{(1)}|_{Y\!uk}$ is the one-loop nonlogarithmic Yukawa contribution
given in the Appendix A.

\section{W-pair production in the electroweak model}
\label{sec4}

\subsection{Analytic results}
\label{sec4.1}

The electroweak Standard Model with the spontaneously
broken $SU_L(2)\times U(1)$ gauge group involves both the massive
$W$ and $Z$-bosons and the massless photon. The corrections to the
fully exclusive cross sections due to the virtual photon exchange
are infrared divergent and should be combined with real photon emission to
obtain infrared finite physical observables.
The infrared divergences of the virtual corrections are regulated by giving
the photon a small mass $\lm$.  In the limit $\lm^2\ll M_W^2\ll Q^2$ the
dependence of the amplitudes on $\lm$ in the full theory is the same as in QED.
Thus the  logarithmic corrections  can be separated into ``pure electroweak''
Sudakov logarithms  and QED Sudakov logarithms of the form $\ln(Q^2/\lm^2)$
or $\ln(M_W^2/\lm^2)$.

To disentangle the electroweak and QED logarithms we use the approach of
Ref.~\cite{Fad,KMPS,JKPS}. While the dependence of the amplitudes on the large
momentum transfer is governed by the {\it hard} evolution equations ({\it  c.f.}
Eqs.~(\ref{evoleqz},~\ref{evoleqa})), their dependence on the photon
mass is governed by the {\it infrared} evolution equations
\cite{Fad}.  Two sets of equations completely determine the dependence of the
amplitudes on two dimensionless variables $Q/M_W$ and $Q/\lm$ up to the initial
conditions which are fixed through the matching to the fixed-order result.
For $\lm^2\ll M_W^2$ the singular dependence of the amplitudes
on the infrared regulator is governed by the QED evolution equation. Its
solution to NNLL accuracy in the massless fermion approximation $m_f=0~(f\ne
t)$ is given by the factor
\bqa
\!\!\!\!
{\cal U}\! &=&
U_0(\al_e)\exp{}\Bigg\{
  \frac{\al_e(\lm^2)}{4\pi}\bigg[
  - \Big( Q_q^2 \!+\! 1 \Big) \ln^2\frac{Q^2}{\lambda^2}
  + \bigg( 3 Q_q^2 \!-\! 4 Q_q \ln\!\frac{\hat{u}}{\hat{t}} \bigg) \ln\frac{Q^2}{\lambda^2}
  + \ln^2\frac{M_W^2}{\lambda^2}
  + 2 \ln\frac{M_W^2}{\lambda^2}
  \bigg]
\nn\\
\!\!
&+&
 \frac{\al_e^2(\lm^{2})}{(4\pi)^2}\,\frac{8}{9}\bigg[\!
  - \!\frac{10}{3}\Big(\! Q_q^2 \!+\! 1 \!\Big)\! \ln^{3}\!\frac{Q^2}{\lambda^2}\!
  + \!\bigg(
          \frac{95}{3}Q_q^2
    \!+\! \frac{50}{3}
    \!-\! 20 Q_q \ln\!\frac{\hat{u}}{\hat{t}}
    \!\bigg)\! \ln^2\!\frac{Q^2}{\lambda^2}\!
  + \!{\cal O}\bigg(\!\! \ln\!\frac{Q^2}{\lambda^2} \!\bigg)
  \bigg]\!
+ {\cal O}(\al_{\!e}^{\!3})
\!\Bigg\},
\label{QED}
\eqa
where
$\al_e$ is the $\overline{\rm MS}$ QED coupling constant, and $Q_q$
is the quark electric charge. The NNLL approximation for ${\cal U}$ can be
obtained from the result for the fermion-antifermion production  \cite{KMPS} by
proper modification of the QED anomalous dimensions. Note that we take into
account the top quark decoupling  and Eq.~(\ref{QED}) corresponds to five
light flavors in contrast to Ref.~\cite{KMPS} where all the quarks were assumed
to be massless. To exclude the top quark contribution in the expressions for the
QED anomalous dimensions in \cite{KMPS} $N_g$ should be replaced by $N_g-1/2$.
The preexponential factor $U_0$ in Eq.~(\ref{QED}) is factorization scheme
dependent. It is convenient to fix it by normalizing
${\cal U}(\al_e)\big|_{s=\lm^2=M_W^2}=1$.
We factorize the QED factor and write the full theory amplitude as a product
\bq
\bfm{\cal A}=
\;{\cal U}\,\bfm{\cal A}_{ew}.
\label{dec}
\eq
where $\bfm{\cal A}_{ew}$ includes only electroweak Sudakov logarithms.
The logarithms of the photon mass in ${\cal U}$ are generated by loops with soft photons,
photons collinear to the initial state fermions, and soft photons collinear
to the final state gauge bosons, which result in the logarithmic dependence of the coefficients on $M_W$.
In the physically motivated cross section which is inclusive in respect to the
photons with the energy much less than electroweak scale the
singular dependence of ${\cal U}$ on the photon mass is replaced by the
experimental cuts on the soft photon energy or absorbed into the parton
distribution functions.
One may easily change the regularization scheme and use {\it e.g.}
dimensional regularization which is more convenient for the analysis of the
parton distribution functions.
In the present paper we focus on the pure electroweak part of the amplitude $\bfm{\cal A}_{ew}$.
The factorization formula~(\ref{dec}) implies that the anomalous dimensions
corresponding to the electroweak Sudakov logarithms are obtained by
subtracting the QED contribution from  anomalous dimensions of the full theory.
The functions  $\gm$, $\zeta$, and $\bfm{\chi}$ are  mass-independent.
Therefore the anomalous dimensions  parametrizing the electroweak logarithms
can be obtained by subtracting the QED contribution from the result of the
unbroken symmetry phase calculation to all orders in the coupling constants.
In particular in one loop we get
\begin{eqnarray}
\gm_{A,\phi}^{(1)} &=&
  \gm_{A,\phi}^{(1)} \Big|_{\ssSU}
- \,{1\over 2}Y_{\!\!A,\phi}^2\,\tw^2
+ \,2\,Q_{\!A,\phi}^2\,\sw^2\,,
\qquad
\zeta_{A,\phi}^{(1)} =
  \zeta_{A,\phi}^{(1)} \Big|_{\ssSU}
+ \,Y_{A,\phi}^2\,\tw^2\,,
\nl
\bfm{\chi}_T^{(1)} &=&
  \bfm{\chi}_T^{(1)} \Big|_{\ssSU}
+ \,4\,Q_q\,\sw^2 \ln\!{u\over t}\,\I,
\qquad\qquad\quad
\bfm{\chi}_L^{(1)} =
  \bfm{\chi}_L^{(1)} \Big|_{\ssSU}
+ \,\Big(
    Y_{\!q}\,Y_{\!\phi}\,\tw^2
  + \,4 \,Q_q\,\sw^2
  \Big)\ln\!{u\over t}\,\I,
\qquad
\label{g1sub}
\end{eqnarray}
where $Y_q$, $Y_{A}= 0$, $Y_{\phi}= -1$ are the hypercharges of quarks,
gauge and Goldstone bosons, $\sw^2=\sin^2\theta_W,~\tw^2=\tan^2\theta_W$,
and $\theta_W$ is the electroweak mixing angle.
The  $SU(2)$ part of the anomalous dimensions can be found in
\cite{Kuhn:2007ca} while the hypercharge contribution and QED subtraction term
are given explicitly.
The anomalous dimensions for the quark ${\cal Z}$-functions can be found in
Ref.~\cite{KMPS}.
The only two-loop coefficients we need are
\begin{eqnarray}
\gm_{A,\phi}^{(2)} &=&
  \gm_{A,\phi}^{(2)}\Big|_{\ssSU}
+ {52\over9} Y_{\!A,\phi}^2\, \tw^4
- {800\over27} Q_{A,\phi}^2\, \sw^4\,,
\label{g2sub}
\end{eqnarray}
in the $\overline{\rm MS}$ scheme.
On the other hand the functions $\xi$ and ${\cal A}_0$ are infrared sensitive
and require the use of the  true mass eigenstates of the Standard Model in the
perturbative calculation.
In NNLL approximation one needs the one-loop contribution to these quantities
which can be found  by comparing the solution of the evolution equation with
the explicit one-loop result for the amplitudes.
In this way we find that the anomalous dimensions $\xi_i^{(1)}$ get contributions just
from the mass difference between $M_W$ and $M_Z$ and obtain:
\bq
\xi_i^{(1)} =
\;2\,\bigg[
  (T_i^3)^2
+ \Big(\frac{Y_i}{2}\Big)^2\,\tw^2
- Q_i^2\,\sw^2
\bigg]\ln\frac{M_Z^2}{M_W^2},
\qquad\qquad
i= \psi,A,\phi,
\eq
where $T_i^3= Q_i - Y_i/2$ is the third component of the isospin.
The expressions for the nonlogarithmic one-loop corrections to the amplitude
${\cal A}_0^{(1)}$ are rather cumbersome and we collect them in Appendix A.
Note that ${\cal A}_0^{(1)}$  depends on the normalization of the QED factor.
We use the normalization where all the nonlogarithmic one-loop corrections are
contained in ${\cal A}_0^{(1)}$.
With the above parameters of the evolution at hand we can write down the
two-loop NNLL corrections to the amplitudes as in Eq.~(\ref{expsola}).
The two-loop Yukawa contribution in the NLL approximation
is given by the interference of the one-loop double logarithms and the
one-loop Yukawa enhanced single logarithms.
Thus it is straightforward to obtain this contribution exactly.
For the NNLL two-loop Yukawa contribution we use the  $SU(2)$ model of
the previous section  with $\rho=m_t^2/(2M_W^2)$, which approximates the
exact result with the accuracy of order $\sin^2\theta_W\approx 0.2$.

Now we are in the position to present the final result for the cross sections.
We define the perturbative series as follows
\begin{equation}
{{\rm d} \sigma \over {\rm d} \cos\theta}=
\left[1+\left( {\alpha \over 4\pi} \right)\dl^{(1)}
+\left( {\alpha\over 4\pi} \right)^2
\dl^{(2)}+\ldots\right]
{{\rm d} \sigma_{LO} \over {\rm d} \cos\theta} \,.
\label{sersig}
\end{equation}
The coefficients for the one and two-loop NNLL terms are listed in the Appendix
B. Below we present the numerical analysis of the corrections to the partonic
and hadronic cross sections.

\subsection{The partonic cross section}
\label{sec41}
\begin{figure}[ht]
  $\!\!\!\!$
  \begin{tabular}{cc}
   \includegraphics[scale=0.64]{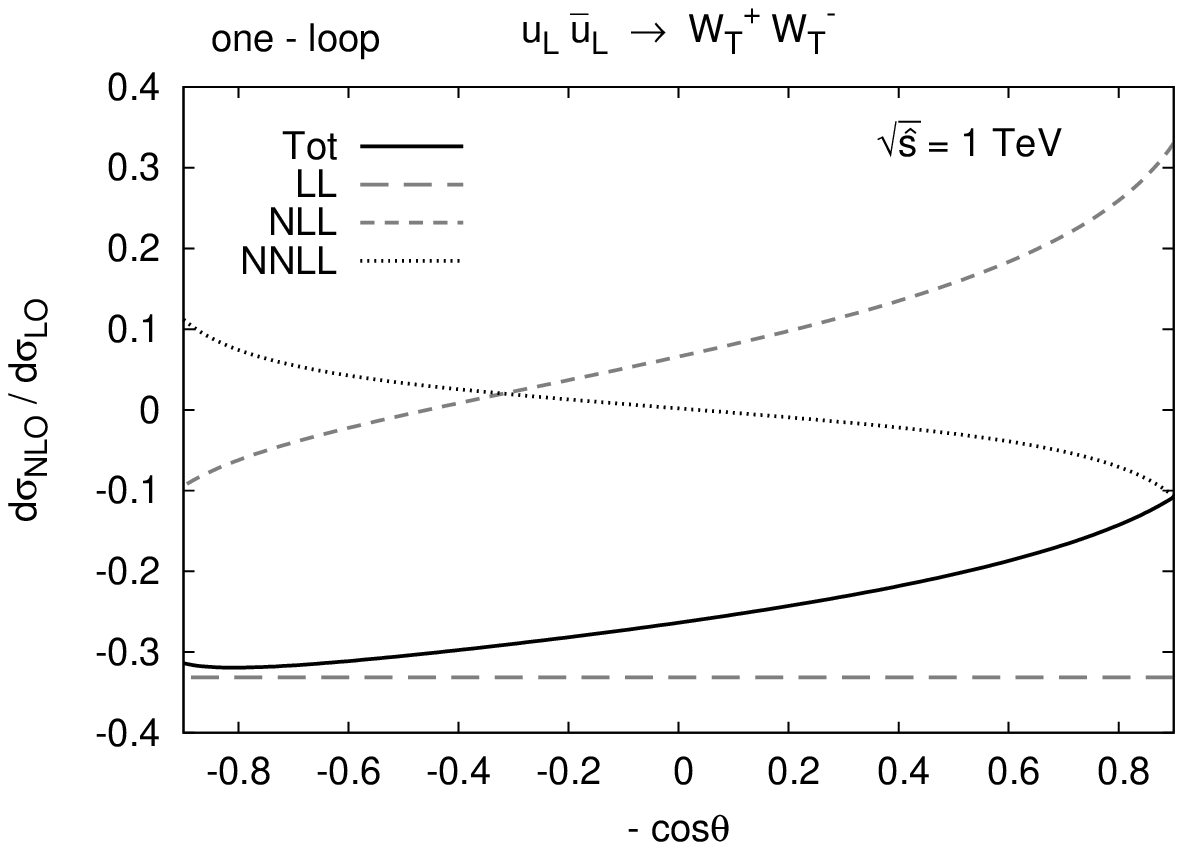}  &
   \includegraphics[scale=0.64]{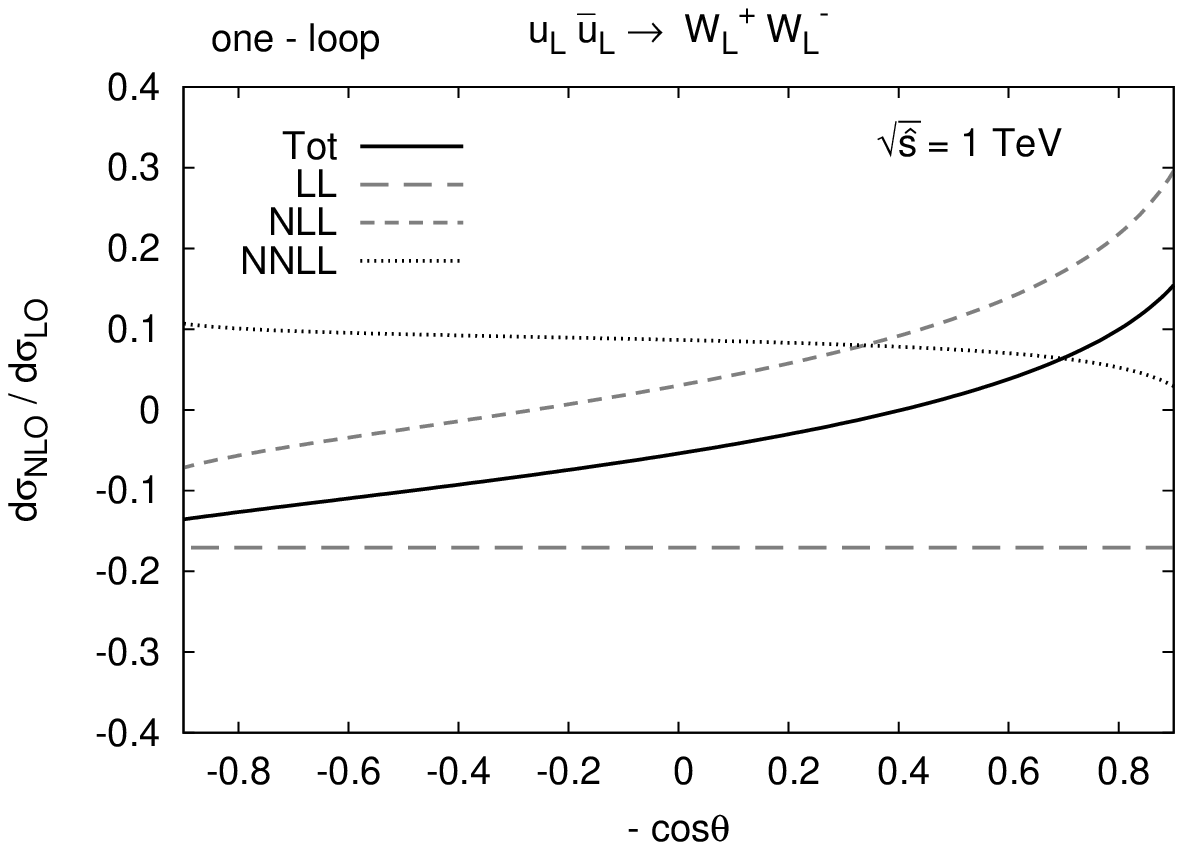}  \\[-0.2cm]
   \includegraphics[scale=0.64]{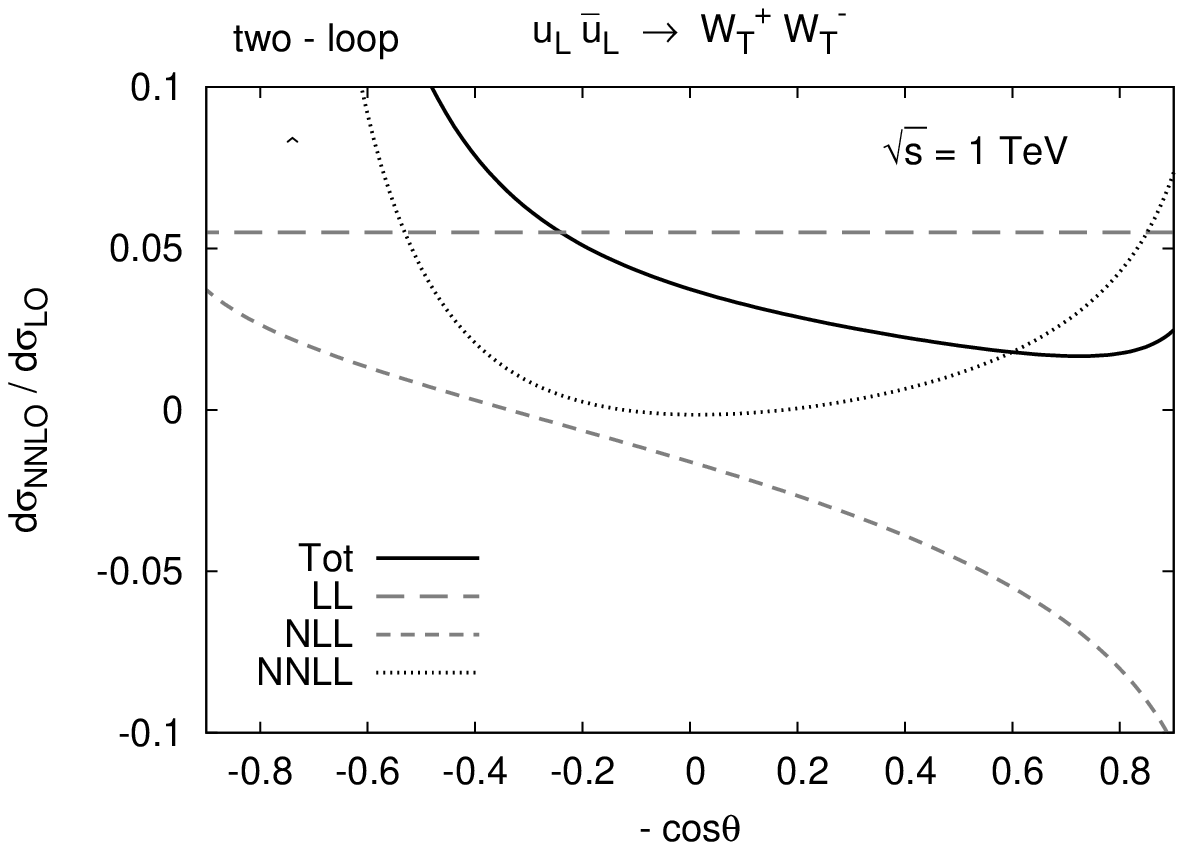} &
   \includegraphics[scale=0.64]{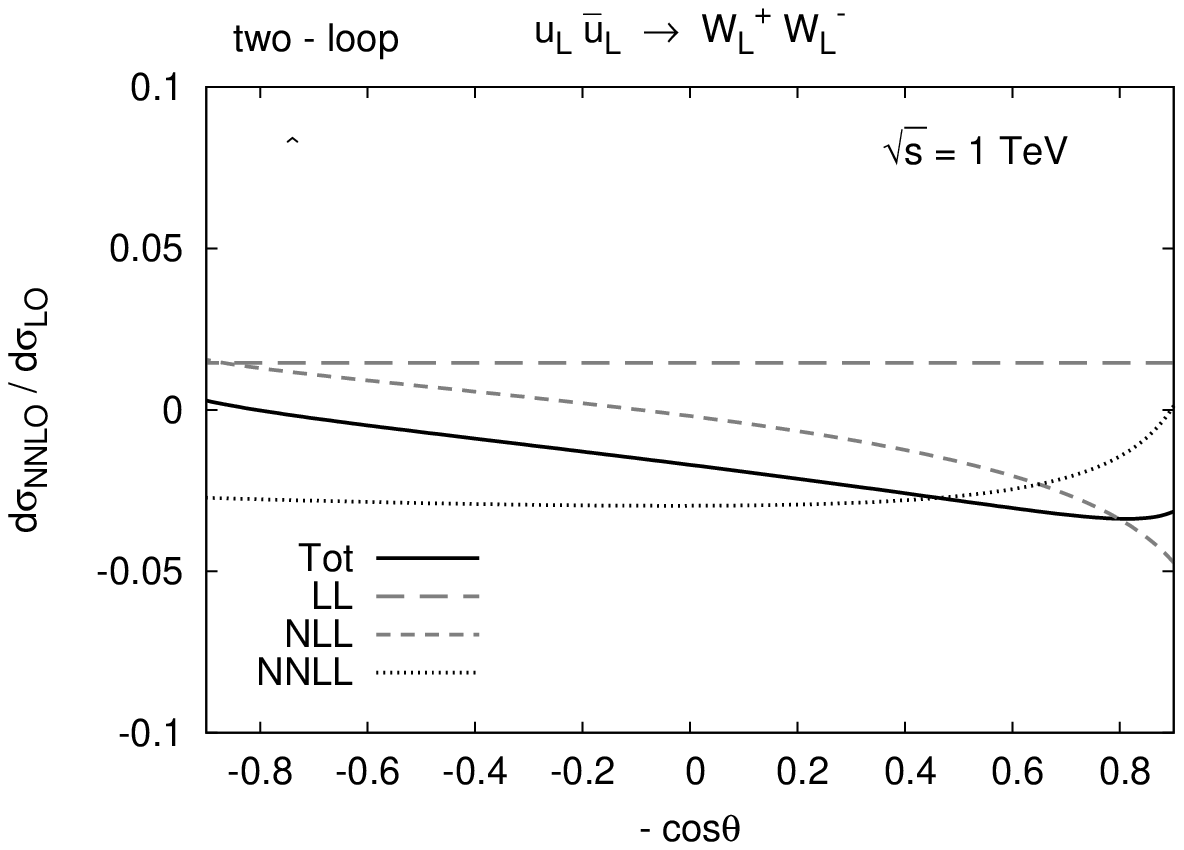} \\[-0.4cm]
  \end{tabular}
  \caption[]
  {One and two-loop corrections to the partonic cross section
   for left-handed $u$-quarks in the initial state,  transverse (left panel)
and
   longitudinal (right panel) $W$-bosons  at $\sqrt{\hat{s}}= 1$ TeV.
  }
  \label{fig:deltaul}
\end{figure}

\begin{figure}[ht]
  $\!\!\!\!$
  \begin{tabular}{cc}
   \includegraphics[scale=0.64]{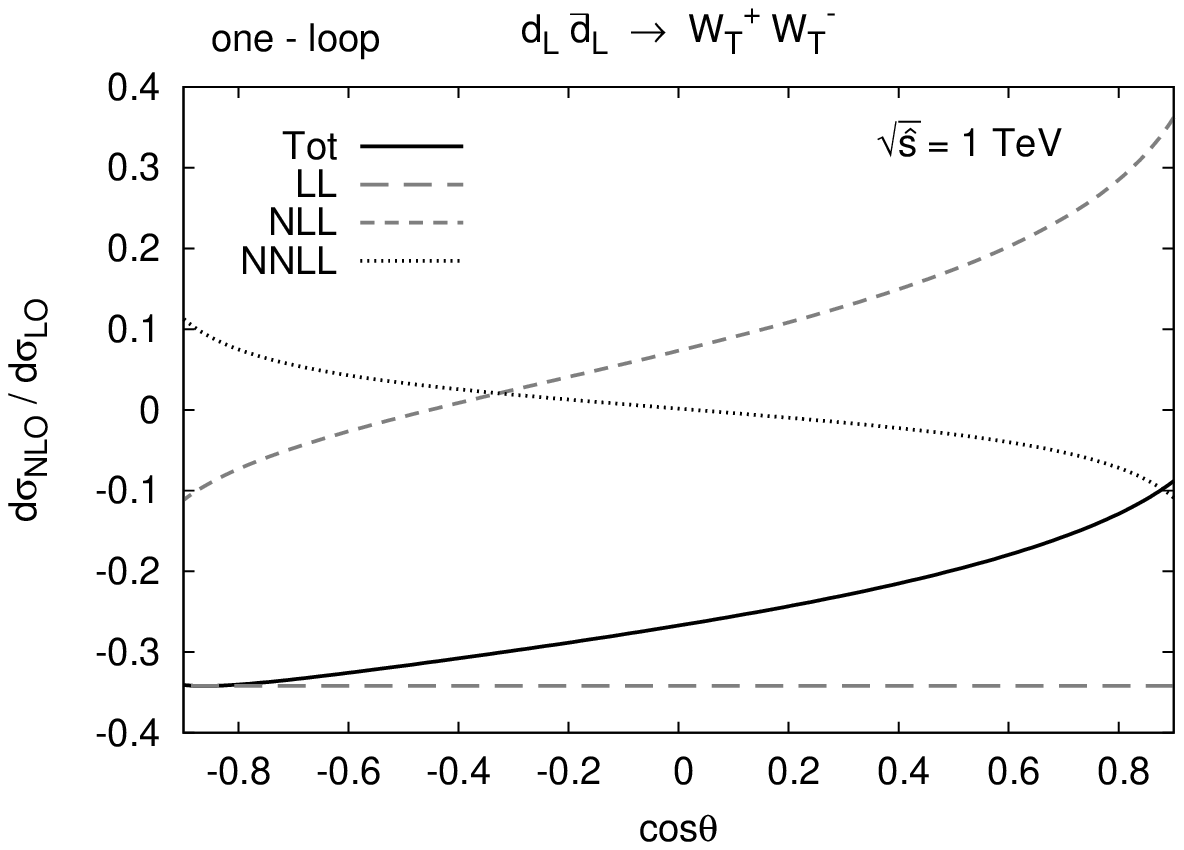}  &
   \includegraphics[scale=0.64]{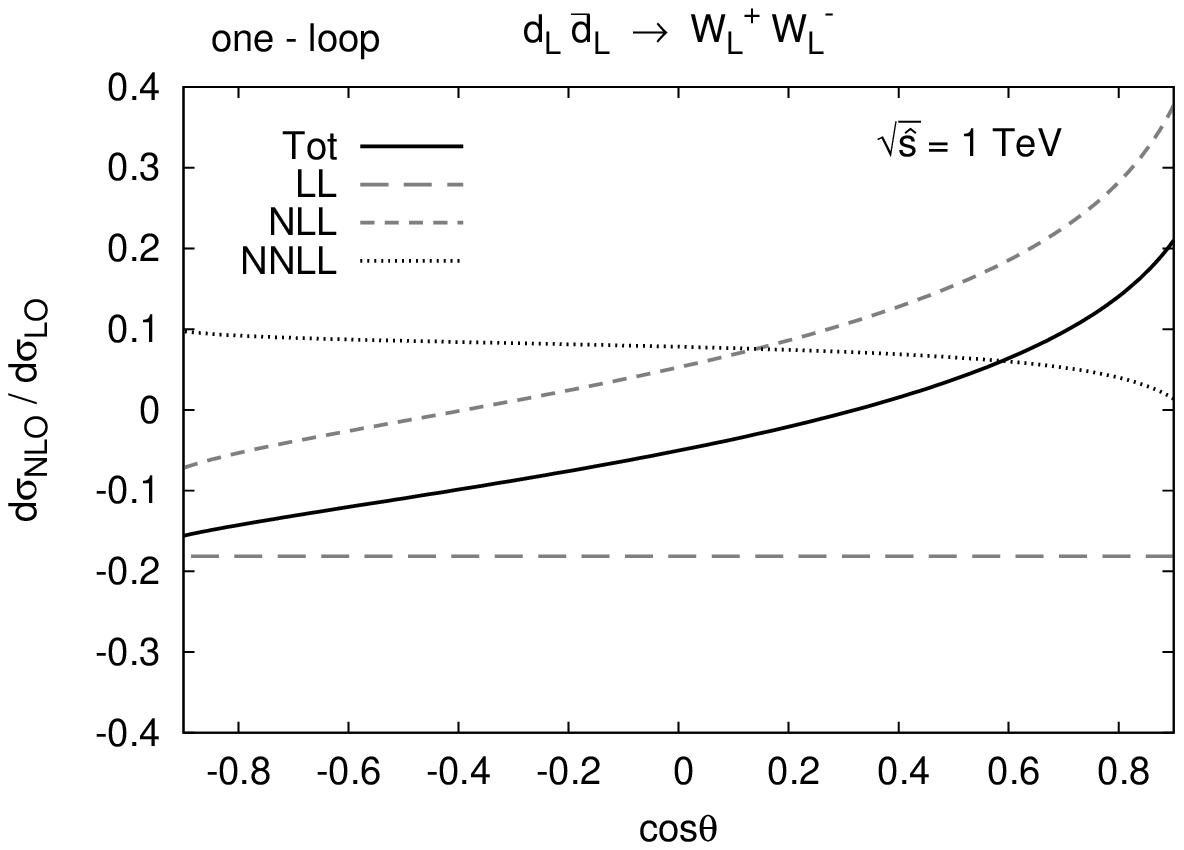}  \\[-0.2cm]
   \includegraphics[scale=0.64]{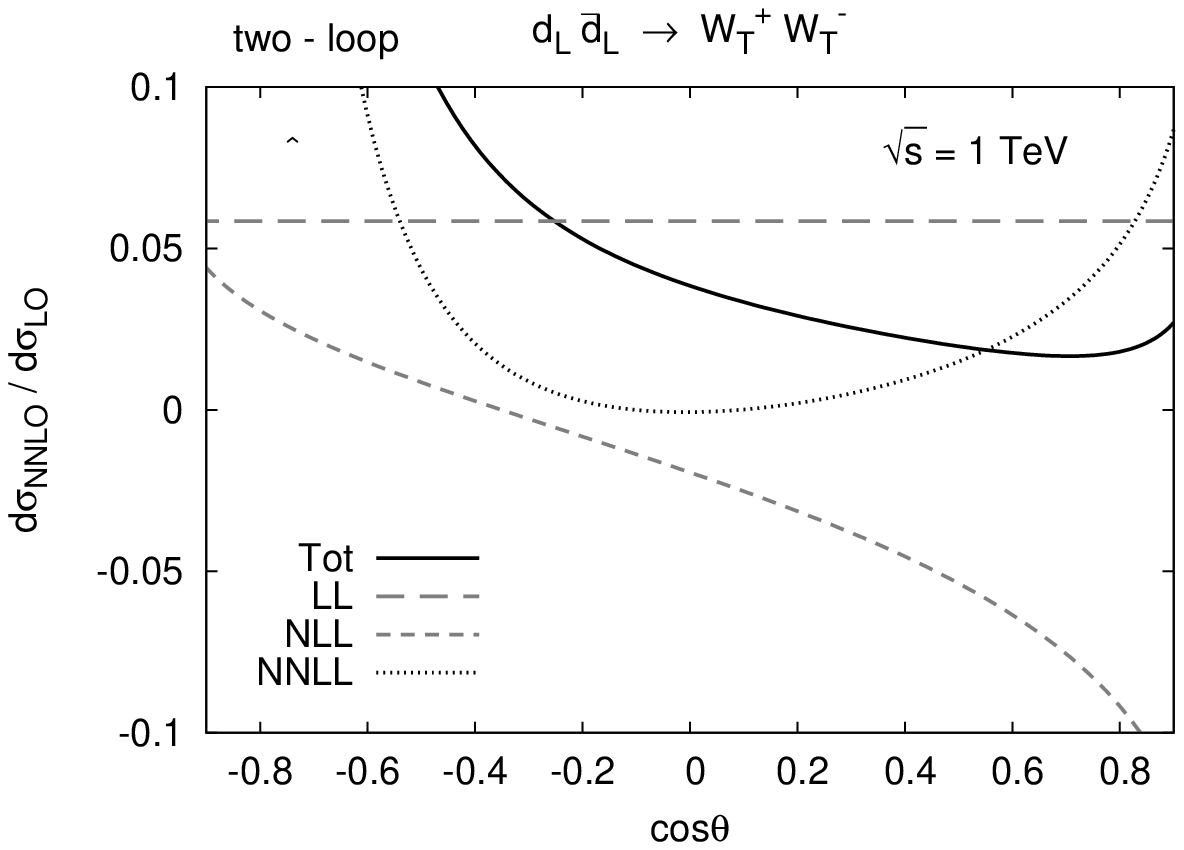} &
   \includegraphics[scale=0.64]{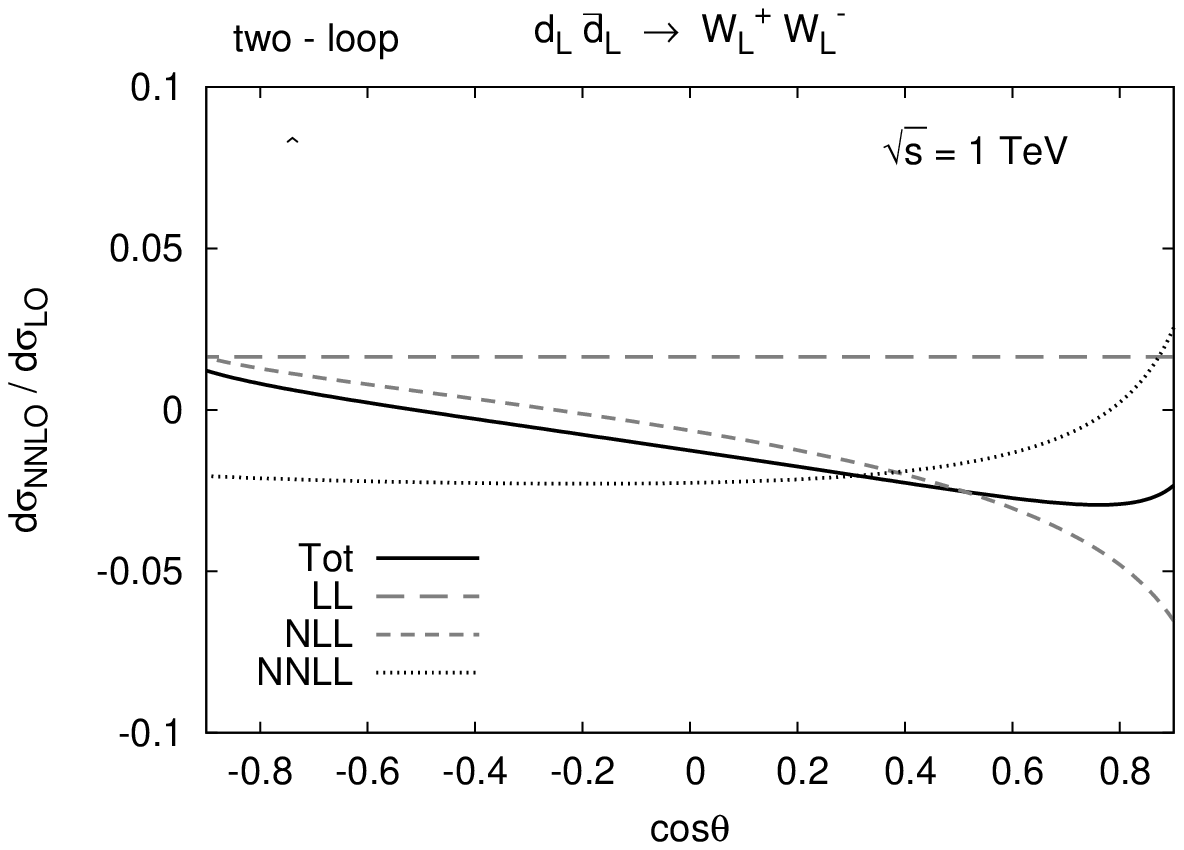} \\[-0.4cm]
  \end{tabular}
  \caption[]
  {One and two-loop corrections to the partonic cross section
   for left-handed $d$-quarks in the initial state, transverse (left panel) and
   longitudinal (right panel) $W$-bosons at $\sqrt{\hat{s}}= 1$ TeV.
  }
  \label{fig:deltadl}
\end{figure}

For the numerical estimates we adopt the following input values
\bq
M_W=80.41\;{\rm GeV},
\qquad
M_Z=91.19\;{\rm GeV},
\qquad
M_H=117\;{\rm GeV},
\qquad
m_t=172.7\;{\rm GeV},
\label{inputparameters}
\eq
$$
\al(M_Z^2)={1 \over 128.1},
\qquad\qquad
s_W^2=0.231,
$$
and  take $\sqrt{s} = 1$~TeV as characteristic example.
The one and two-loop corrections for left-handed $u$-quarks in the initial
state are plotted in \fig{fig:deltaul} showing a sizable NNLL
contribution\footnote{
Numerical results for the partonic cross section have been presented in
Ref.~\cite{Man2} and qualitatively agree with our analysis. However a direct
comparison of the results is not possible since the authors of \cite{Man2} use
different power counting and QED subtraction prescription.
}. The structure of the corrections for the left-handed $d$-quarks is similar, see
\fig{fig:deltadl}. To facilitate the comparison of the  $u$- and   $d$-quarks cases
related by crossing symmetry in the Born approximation, we plot the cross section for 
$u$-quarks as a function of $-\cos\theta$. In the Born cross section we 
always use the physically motivated normalizarion scale of the coupling constants, 
which is  $\mu=M_W$  for the transverse and  $\mu=\sqrt{s}$ for the longitudinal boson 
production.

\begin{figure}[ht]
  $\!\!\!\!$
  \begin{tabular}{cc}
   \includegraphics[scale=0.64]{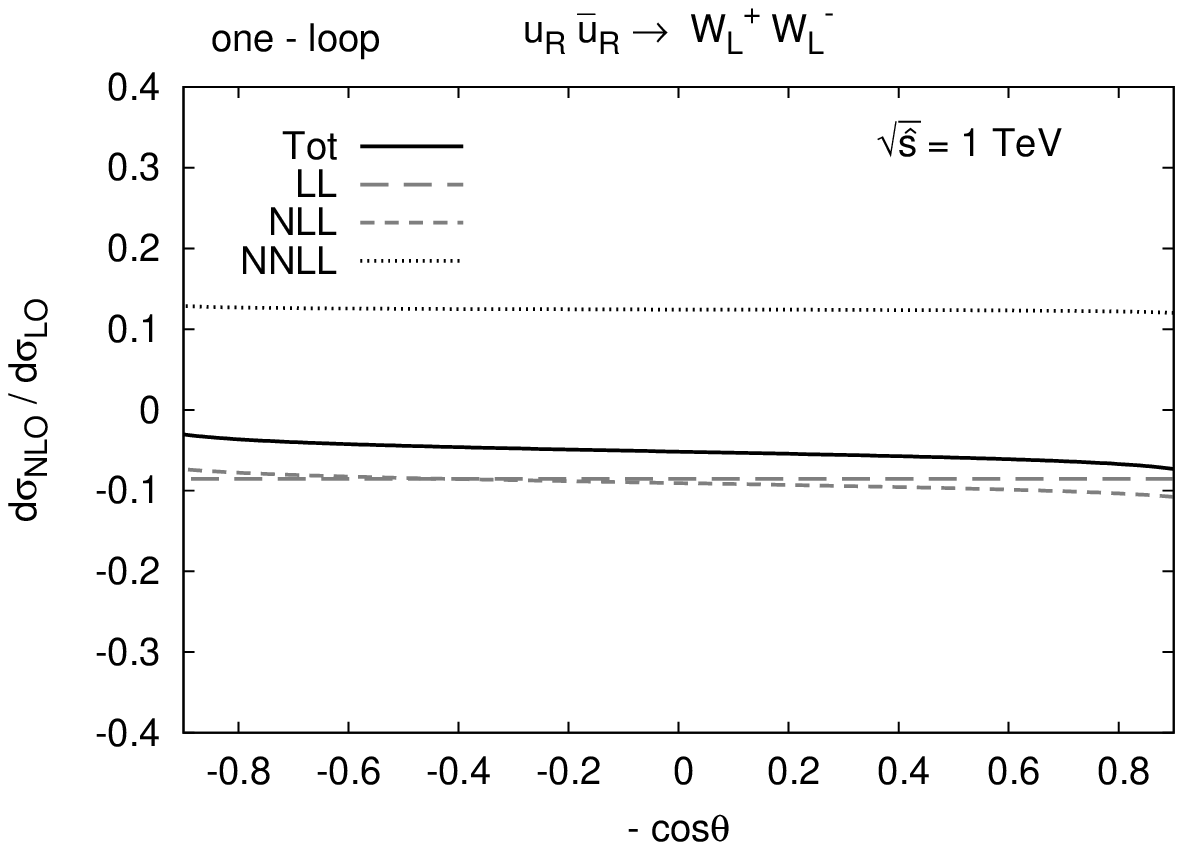}  &
   \includegraphics[scale=0.64]{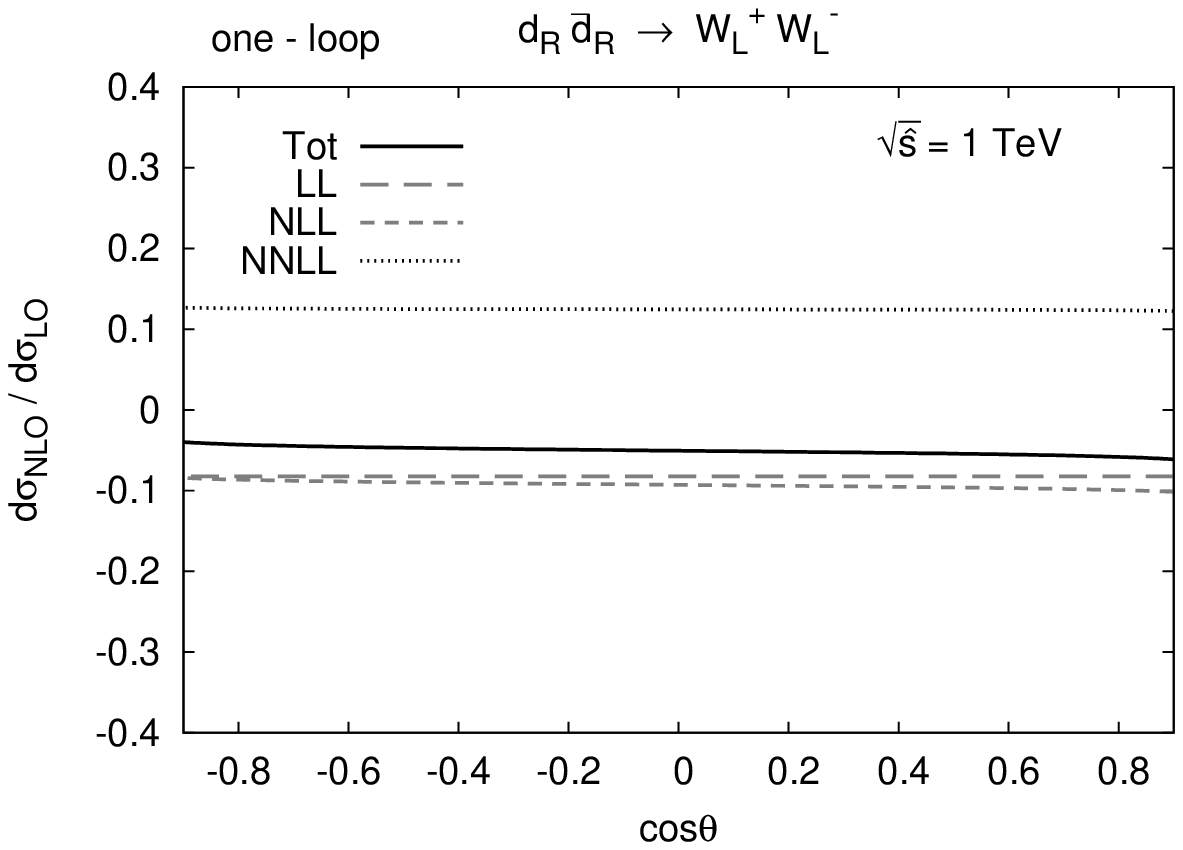}  \\[-0.2cm]
   \includegraphics[scale=0.64]{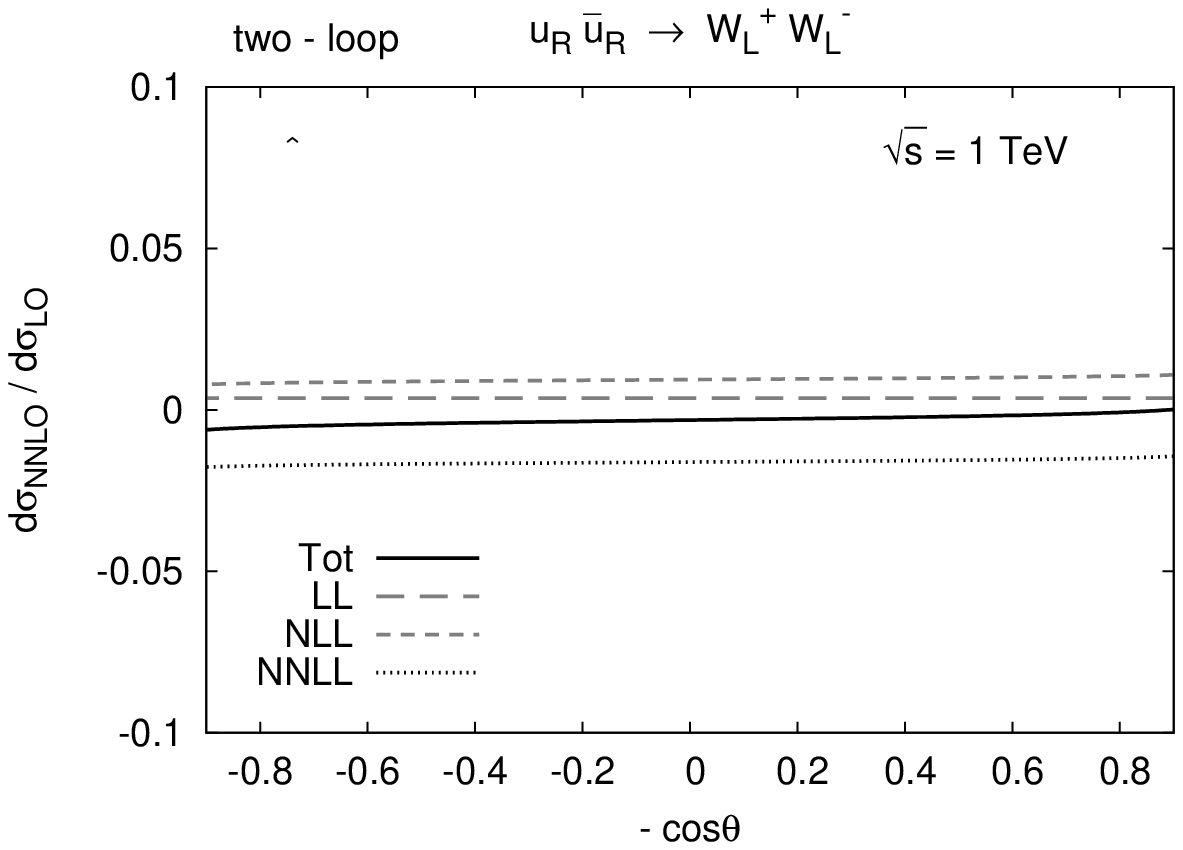} &
   \includegraphics[scale=0.64]{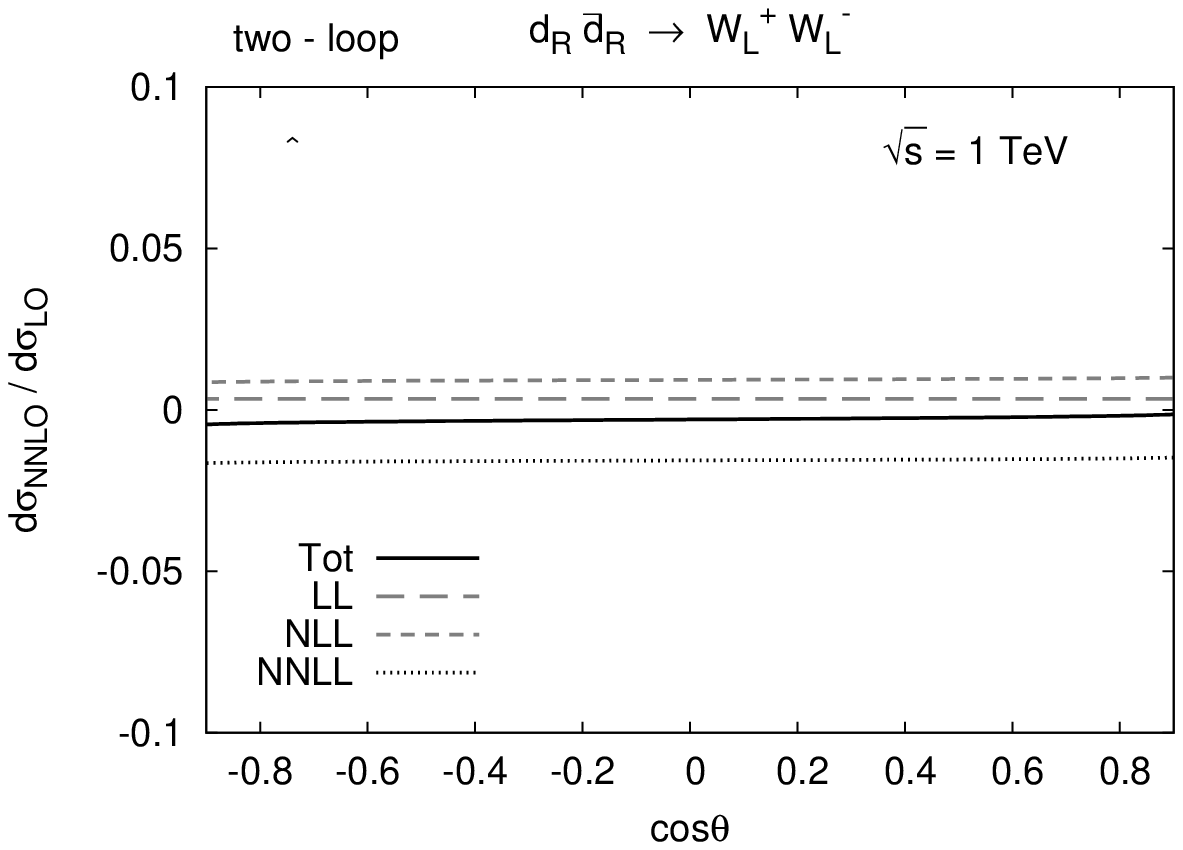} \\[-0.4cm]
  \end{tabular}
  \caption[]
  {One and two-loop corrections to the partonic cross section
   for right-handed $u$-quarks (left panel) and $d$-quarks  (right panel) in the
   initial state,   and longitudinal $W$-bosons at  $\sqrt{\hat{s}}= 1$ TeV.
  }
  \label{fig:deltar}
\end{figure}

The contribution of the  right-handed quarks vanishes for
transversally polarized $W$-bosons, and for longitudinally polarized
bosons it  is significantly smaller than the one of left-handed quarks, see
\fig{fig:deltar}.
In one as well as in two-loop approximation one observes large compensations
between LL, NLL and NNLL terms.
Evidently the LL approximation, even when combined with NLL terms only, does
not lead to an adequate description of the full result.
In Ref.~\cite{Bee} the quality of the high energy approximation has been
studied at one-loop level. The error turns out to be less than a few percents for
a partonic center of mass energy above $500$ GeV and a scattering angle
in the range $30^o<\theta<150^o$.

\subsection{Hadronic cross section}
\label{sec42}
\begin{figure}[ht]
 \begin{center}
   \includegraphics[scale=0.62]{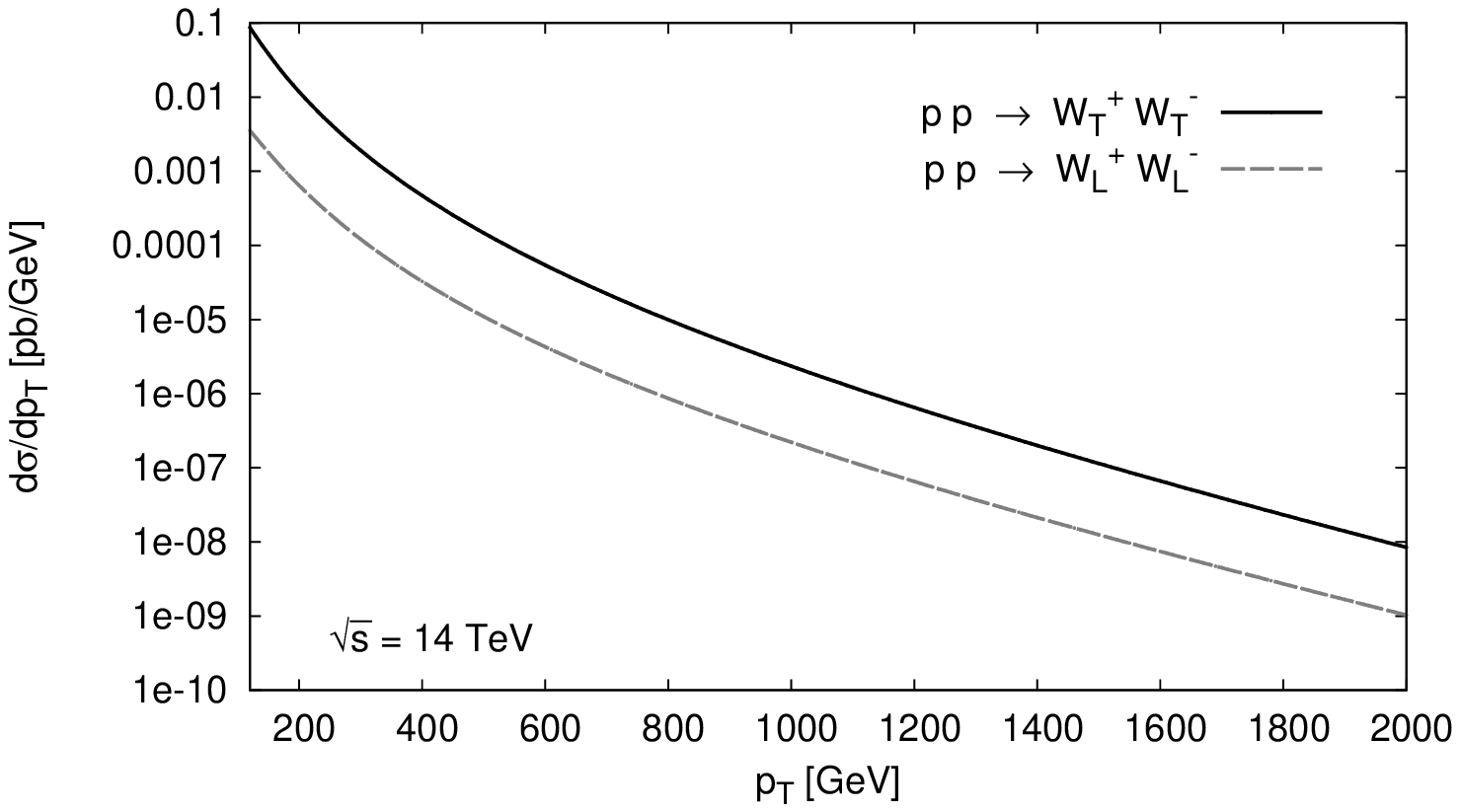}   \\
  \begin{tabular}{cc}
   \includegraphics[scale=0.62]{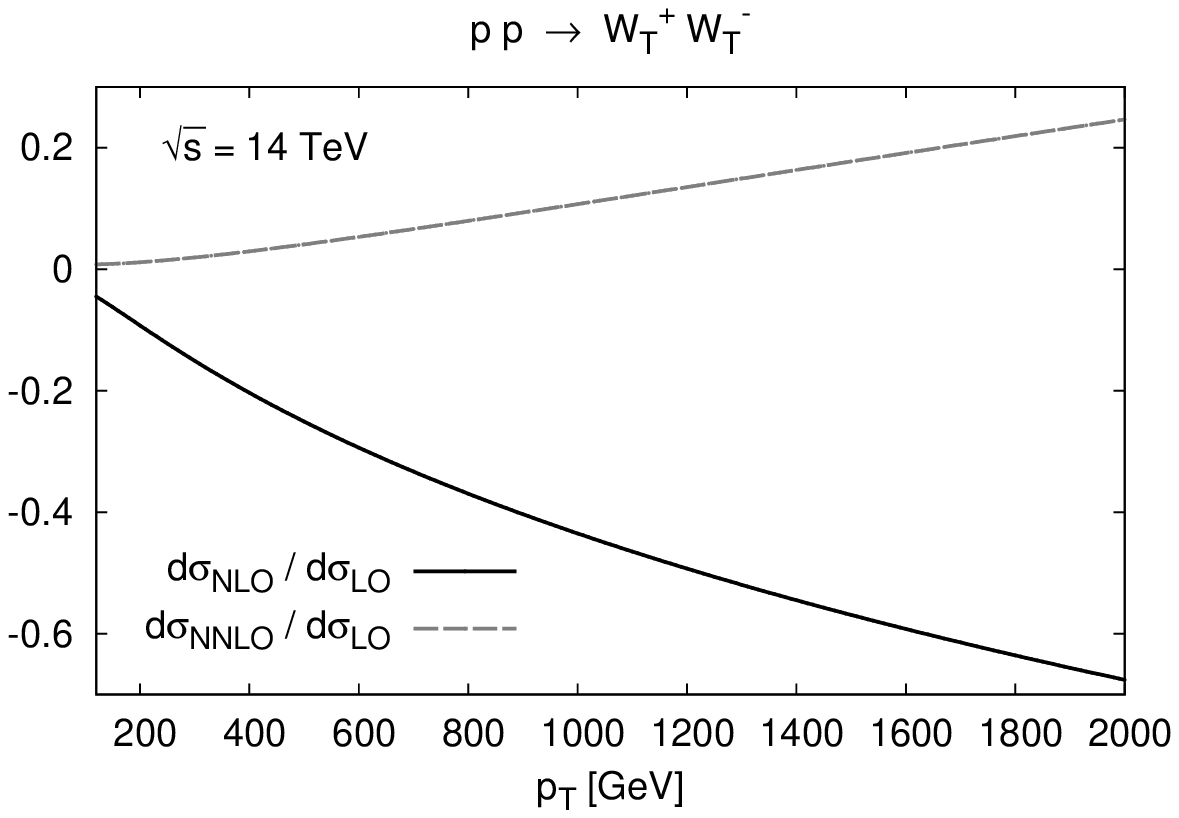}    &
   \includegraphics[scale=0.62]{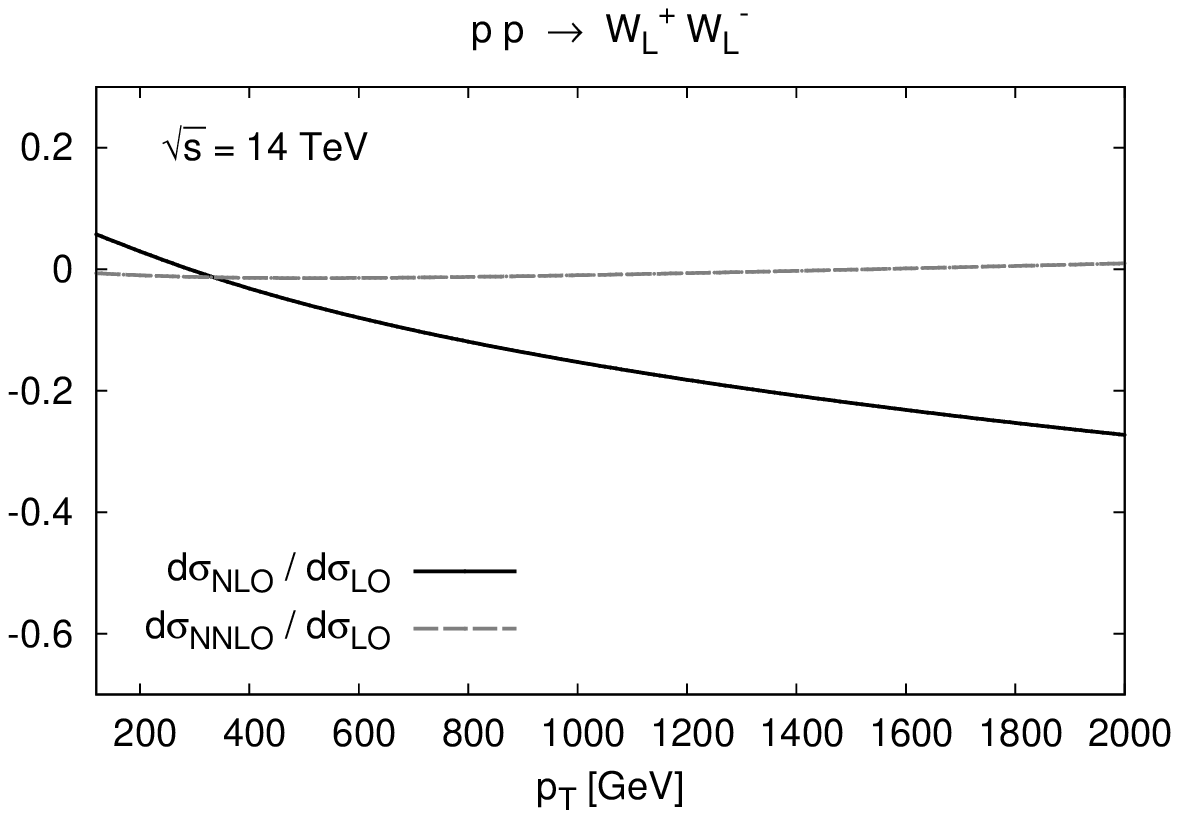}
  \end{tabular}
  \caption[]{
  Transverse momentum distribution (including corrections) of transverse and longitudinal W pairs and
  relative corrections for proton-proton collisions at $\sqrt{s}= 14$ TeV.
  }
  \label{fig:pt}
 \end{center}
\vspace{-0.5cm}
\end{figure}

To obtain transverse momentum and invariant mass distributions for the process
$pp\to W^+W^- +X$ the partonic cross section must be convoluted with the parton distribution
functions $f_{h_1,i}(x_1,\mu_F^2)$ and $f_{h_2,j}(x_2,\mu_F^2)$,
where $\mu_F$ is the factorization scale, $x_1$ and $x_2$ are the momentum
fractions carried by the parton $i$ in the hadron $h_1$ and by the parton $j$ in
the hadron $h_2$ respectively. The $p_T$-distribution is given by
\bq
{d\sigma\over dp_T}=
{1\over N_c^2}\,\sum_{ij}
\int_0^1\!\!dx_1\int_0^1\!\!dx_2\,
f_{h_1,i}(x_1,\mu_F^2)\,f_{h_2,j}(x_2,\mu_F^2)\,
\theta(x_1x_2-\tau_{\rm min})\,
{d\hat{\sigma}_{ij}\over dp_T}\,,
\eq
where $N_c$ is the number of colors, the sum is over all possible $q\bar{q}$
partonic initial state, $\pt=\sin\theta\sqrt{\shat-4M_W^2}\Big/2$ is the
transverse momentum of the $W$-bosons and we adopt $\mu_F=\pt$.
The quantity
\bq
\tau_{\rm min}={4(p_T^2+M_W^2)\over s}
\eq
is related to the minimal partonic energy that is
needed to produce two $W$-bosons with a given transverse momentum $\pt$.
The partonic differential cross section $d\hat{\sigma}_{ij}/dp_T$ are given
in terms of the  angular differential cross section as follows
\bq
{d\hat{\sigma}_{ij}\over dp_T}=
{4p_T\over \sqrt{\hat{s}-4M_W^2}\sqrt{\hat{s}-s\,\tau_{\rm min}}}\,
\Big[\,
  {d\hat{\sigma}_{ij}\over d\cos\theta} + (\hat{t}\leftrightarrow\hat{u})
\,\Big],
\qquad
\hat{s}= x_1\,x_2\,s\,.
\eq
The numerical results are obtained by using the MRST parton distributions
\cite{Martin:2002dr} and the integration routine {\it CUHRE} from the
{\it CUBA} library \cite{cuba}.
The upper panel of \fig{fig:pt} shows the NNLO $\pt$-distributions
for the production of transverse and longitudinal $W$-bosons  in the NNLL
approximation. Transverse bosons production is  evidently dominant, with the
cross section being about twenty times larger than the one of the  longitudinal
bosons. The lower panel of \fig{fig:pt} shows the NLO and NNLO
corrections  separately.
For the production of transversely polarized $W$-pairs the one-loop correction
reaches  40\% at $\pt=1$~TeV and 60\% at $\pt=2$~TeV.
The two-loop contribution amounts up to 10\%  at $\pt=1$~TeV and 20\%
at $\pt=2$~TeV and  partially compensate  the one-loop
corrections.  For the longitudinal boson production the one-loop correction is
about 15\% (30\%) at  $\pt=$1~TeV ($\pt=$2~TeV), while the two-loop contribution
does not exceed  a few percent up to $\pt=2$~TeV.
As anticipated above the radiative corrections for the longitudinal case are
smaller than those for transverse $W$ bosons.
This is because the value of the quadratic Casimir operator of the $SU_L(2)$ electroweak group,
which govern the leading logarithmic contribution, is smaller for the fundamental representation
of the longitudinal degrees of freedom than for the adjoint representation of the
transversely polarized $W$-bosons.

\begin{figure}[h]
 \begin{center}
   \includegraphics[scale=0.62]{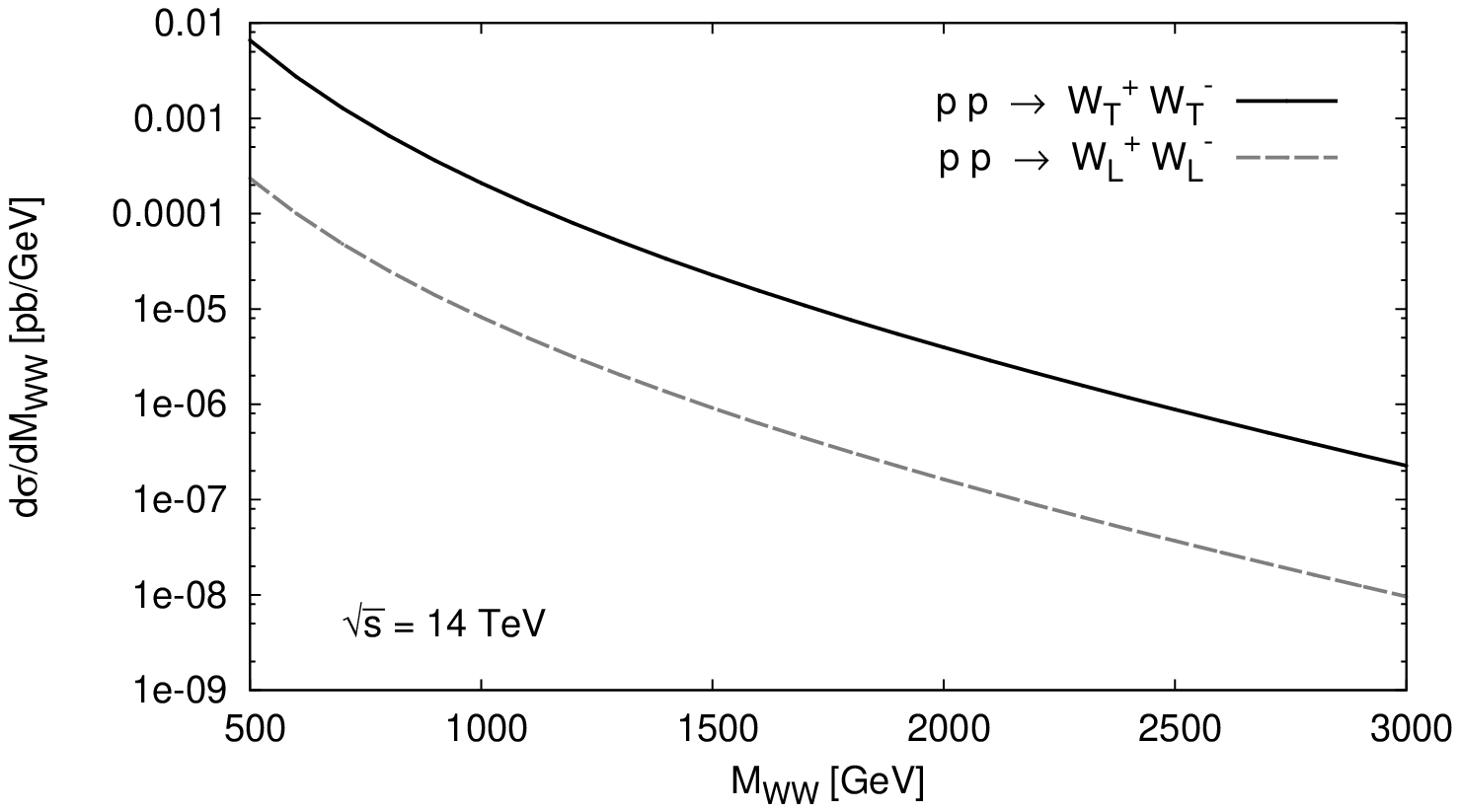}   \\
  \begin{tabular}{cc}
   \includegraphics[scale=0.62]{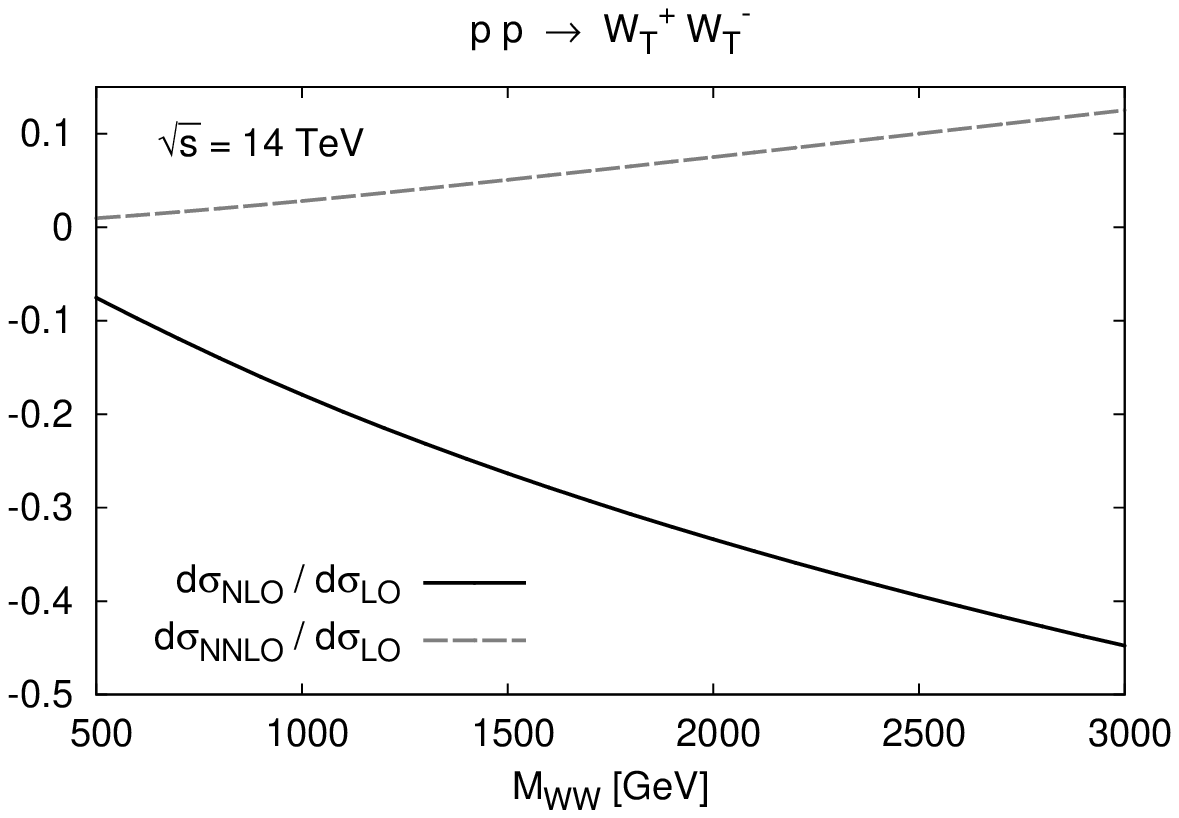}    &
   \includegraphics[scale=0.62]{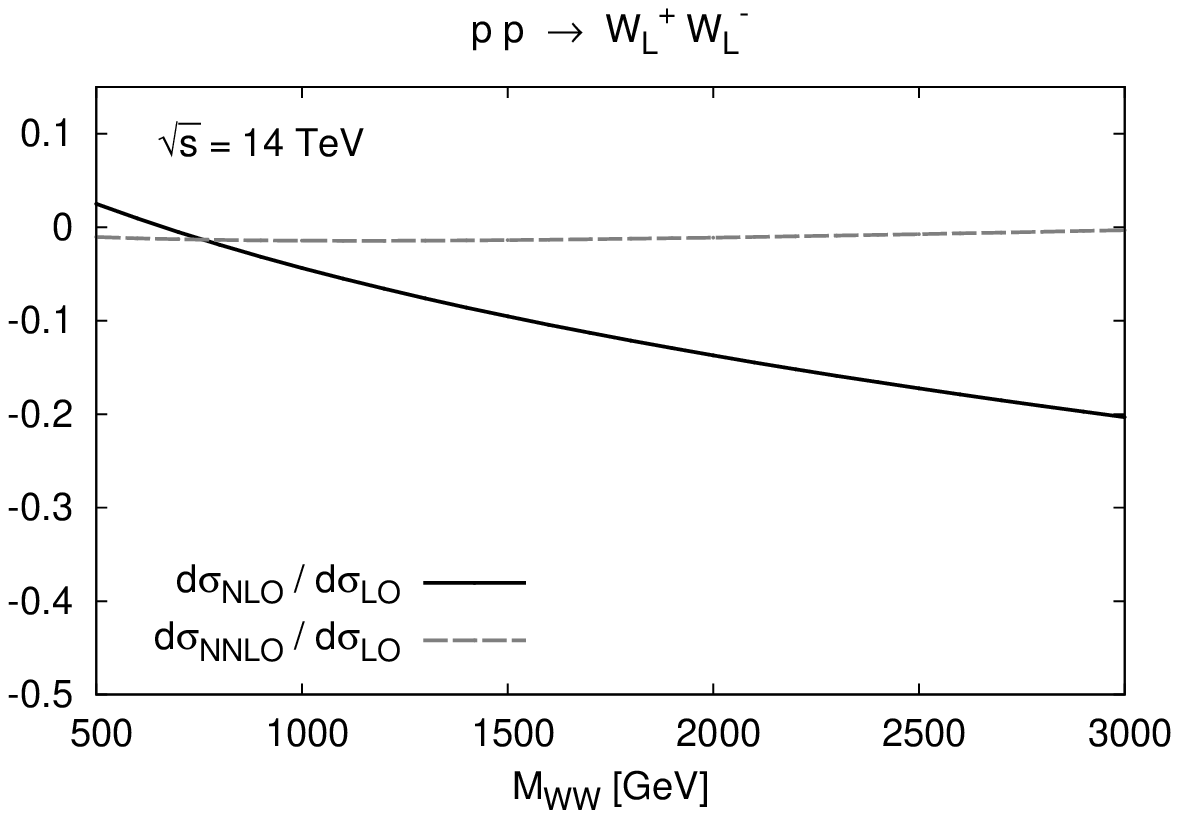}
  \end{tabular}
  \caption[]{
  Total invariant mass distribution (including corrections) of transverse and longitudinal $W$-pairs
(upper panel) and  the corrections  to the invariant mass distribution (lower
panel) for proton-proton collisions at $\sqrt{s}= 14$ TeV.
  }
  \label{fig:Mww}
 \end{center}
\vspace{-0.5cm}
\end{figure}

The invariant mass distribution for the $W$-pair production is defined as
follows
\bq
{d\sigma\over d M_{_{WW}}}=
{1\over N_c^2}\sum_{ij}\int_0^1dx_1\int_0^1dx_2
f_{h_1,i}(x_1,\mu_F^2)f_{h_2,j}(x_2,\mu_F^2)
{{\rm d}\sigmahat_{ij}(M_{_{WW}}^2,\alpha)\over{\rm d}M_{_{WW}}}\,,
\eq
where $M_{_{WW}}=\sqrt{(k_++k_-)^2}=\sqrt{\hat{s}}$ is the invariant mass
of the $W$-pair system and we adopt  $\mu_F=M_{_{WW}}$.
Here the partonic differential cross section
${\rm d}\sigmahat_{ij}/{\rm d}M_{_{WW}}$ is obtained by integrating
the angular differential cross section in the region
$-\cos\theta_{\rm min}<\cos\theta<\cos\theta_{\rm min}$
\bq
{{\rm d}\sigmahat_{ij}(M_{_{WW}}^2)\over{\rm d}M_{_{WW}}}=
\int_{-\cos\theta_{\rm min}}^{\cos\theta_{\rm min}}{\rm d}\cos\theta
{{\rm d}\sigmahat_{ij}(M_{_{WW}}^2)\over{\rm d}\cos\theta}
\delta(\sqrt{x_1x_2s}-M_{_{WW}})\,,
\eq
which excludes the range of small angles where the high energy and the Sudakov
approximations are not valid. The results for the invariant mass distribution
are plotted in \fig{fig:Mww} with an angular cutoff
$\theta_{\rm min}=30^o$.
To estimate the potential statistical sensitivity, the corresponding plots are
shown for the production cross section of W pairs with $p_T\ge p_T^{\rm cut}$.
Taking, as crude estimate, an integrated luminosity of 200 fb$^{-1}$, about 1200
W pairs with $p_T>600$ GeV would be produced. Assuming that the experimental analysis
would be based on the final state with one $W$-boson decaying leptonically
and the other hadronically  a fraction of about 4/9 of the pairs could be observed, 
leading to a nominal statistical error of about 4\%.
Under this (optimistic) assumption the one-loop terms would be clearly relevant
and the two-loop terms start to contribute.
\begin{figure}[t]
 \begin{center}
   \includegraphics[scale=0.62]{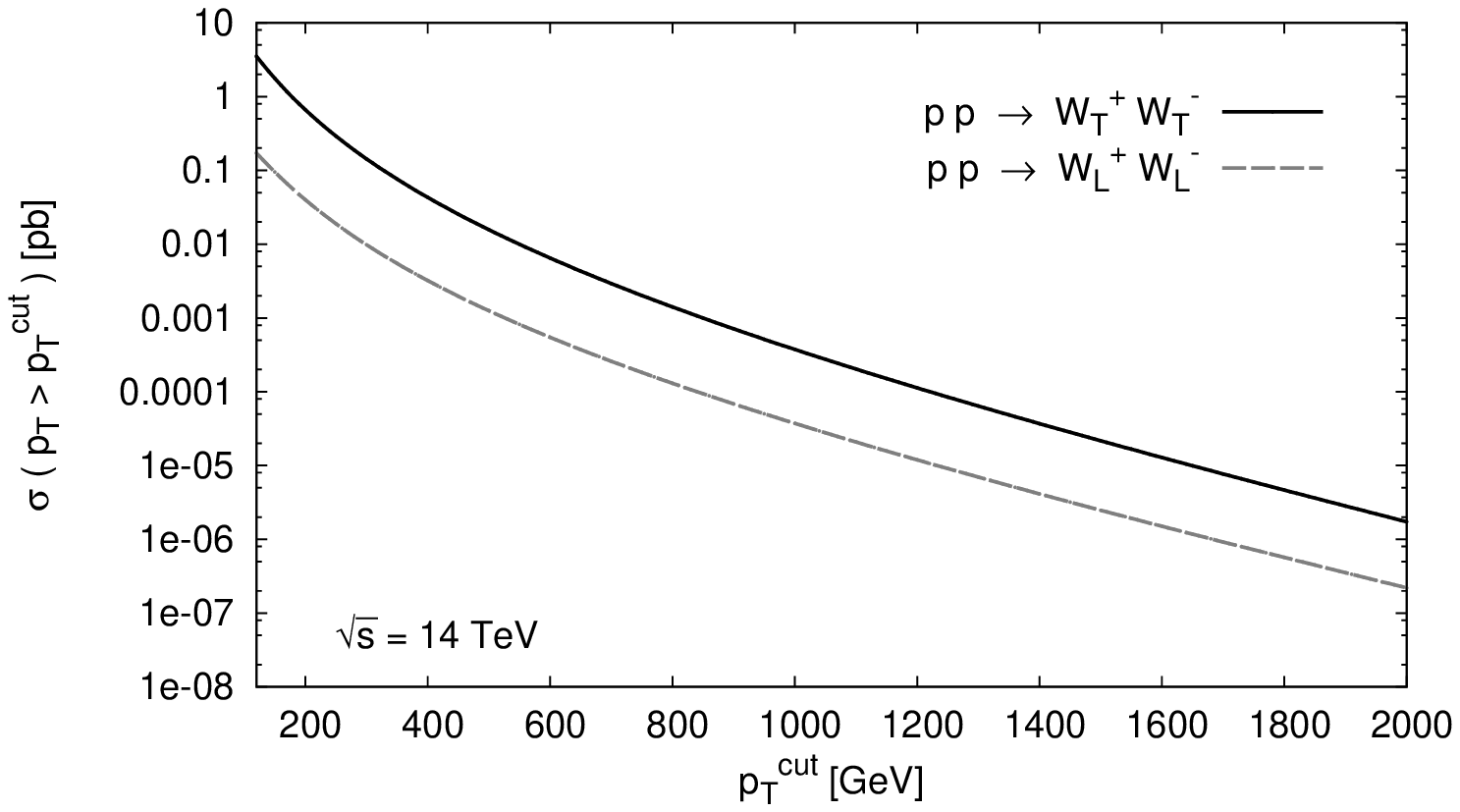}   \\
  \begin{tabular}{cc}
   \includegraphics[scale=0.62]{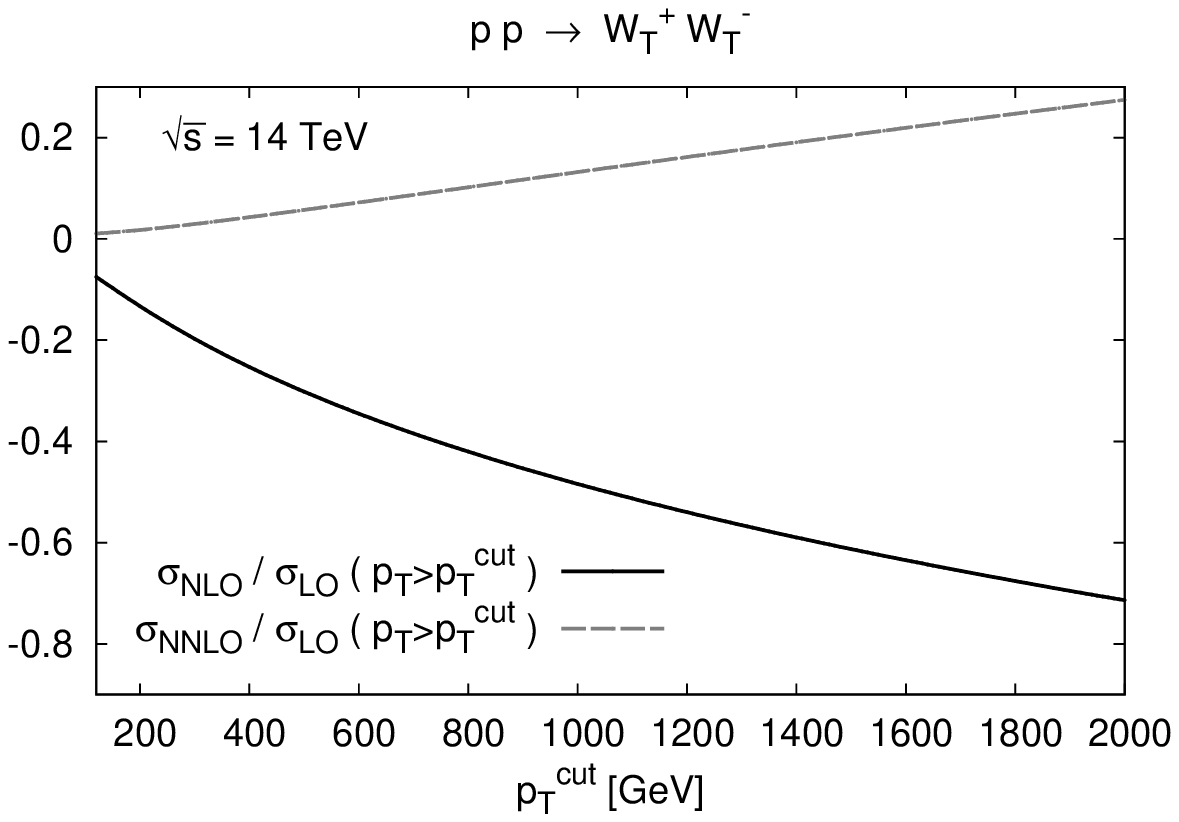}    &
   \includegraphics[scale=0.62]{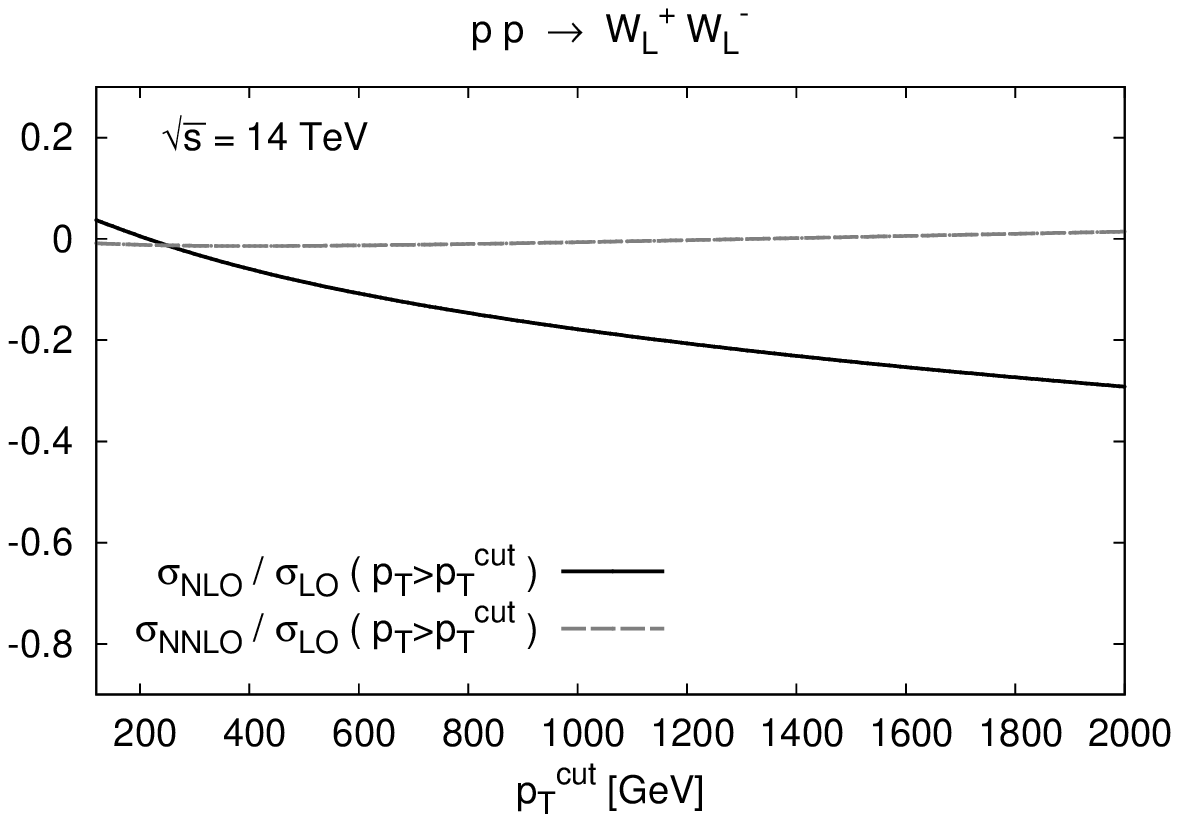}
  \end{tabular}
  \caption[]{
  Cross section (including corrections) for transverse and longitudinal W pair production and
  relative corrections for proton-proton collisions at $\sqrt{s}= 14$ TeV (see text).
  }
  \label{fig:pt0}
 \end{center}
\vspace{-0.5cm}
\end{figure}

\section{Summary}
\label{sec5}
In the present paper we derived the one and two-loop electroweak corrections
to $W$-pair production at the LHC in NNLL approximation in high energy limit.
We present the analytical result for the amplitudes,
differential partonic cross sections, hadronic $p_T$- and invariant mass
distributions. The structure of the corrections  is similar to the
$W$-pair production in $e^+e^-$ annihilation \cite{Kuhn:2007ca}.
In the  case of the transverse boson production we observe the cancellation
between the huge NLL and NNLL contributions so that
the sum is dominated by the LL term.
For the longitudinal bosons the corrections exhibit significant cancellation
between the LL, NLL and NNLL terms.
The maximal effect of the corrections is on the $p_T$-distribution of the
transverse $W$-pair production and reaches 60\% and 20\% at $p_T=2$~TeV in one
and two loops, respectively.
To push the theoretical error below 1\% the evaluation of the two-loop linear
logarithmic terms should be completed, which requires the calculation of the
two-loop mass-dependent anomalous dimensions \cite{JKPS}.

\Acknowledgments
The work of A.P. is supported by NSERC, Alberta Ingenuity foundation, and DFG Mercator grant.
The work of S.U. is supported  by BMBF contract 05HT4VKAI3 and GRK 742 ``High
Energy Physics and Particle Astrophysics''.

\section*{Appendix A}

In this appendix we give the explicit analytical result, valid in the high
energy limit, for the one and two-loop NNLL amplitudes  of the processes:
\bqa
A_{\ssT q_-}:\quad
q_-(p_1) + \bar{q}_+(p_2) &\to& W_T^+(k_+) + W_T^-(k_-)\,,
\nn\\
A_{\ssL q_\mp}:\quad
q_\mp(p_1) + \bar{q}_\pm(p_2) &\to& \phi^+(k_+) + \phi^-(k_-)\,,
\eqa
where $q_\mp$ are the left/right handed fermions in the initial state, which
can be either $u$ or $d$ quarks.
The amplitude $A_{T}$ describes the production of transversely polarized
$W$-bosons and vanishes for the right-handed initial quarks.
The amplitude $A_{\ssL\mp}$ describes through the one-loop Goldstone
equivalence theorem the production of the longitudinally polarized $W$-bosons.
The one-loop corrections to the Goldstone equivalence theorem of
\fig{fig:goldstone} can be properly described by introducing the following 
effective wave function counterterm for the $\phi^\pm$ field:
\bq
Z_\phi^{1/2} =
  1
- {\Sigma_L^W(M_W^2)\over M_W^2}
- {\Sigma^{W\phi}(M_W^2)\over M_W}
+ {1\over 2}{\dl M_W^2\over M_W^2}
+ {1\over 2}\dl Z_W
+ {\cal O}(\al^2)\,,
\eq
where $\Sigma_L^W$ is the longitudinal part of the $W$ self-energy,
$\Sigma^{W\phi}$ is the $W\!\!-\!\phi$ self-energy, $\dl M_W^2$ and $\dl Z_W$ are the
mass and wave function counterterms of the $W$ boson (see also 
\cite{BDJ}).

The results given in this section are obtained by adopting the
$\overline{\rm MS}$ renormalization for the couplings and the weak mixing
angle and on-shell renormalization for the masses.
As before the renormalization scale in the Born amplitudes is fixed to $\mu^2_T=M_W^2$ for the transverse and
$\mu^2_L=s$ for the longitudinal case.
The Lorentz-Dirac structure of the amplitudes in the high energy limit takes
a simple form:
\bqa
A_{\ssT q_-}
&=&
\bar{q}(p_2)\,\Big[\,
  \Slash{\epsilon}_\kappa^*\,p_1\!\cdot\!\epsilon_\kappa^*\,
  {\cal A}_{\ssT q_-}\,
+ \,\Slash{k_+}\,(p_1\!\cdot\!\epsilon_\kappa^*)^2\,
  {\cal B}_{\ssT q_-}
\,\Big]\,\omega_-\,{q}(p_1),
\nn\\
A_{\ssL q_\mp}
&=&
\bar{q}(p_2)\,\Slash{k_+}\,\omega_\mp\,{q}(p_1)
\,{\cal A}_{\ssL q_\mp},
\label{dirac-structure}
\eqa
where $\omega_\pm={1\pm\gamma^5 \over 2}$ and we use the relation between the
polarization vectors $\epsilon^\mu_{\kappa}(k_\pm)$ of the transversely polarized $W^\pm$
in the center of mass frame
\bq
\epsilon^\mu_{\kappa}(k_+)=-\epsilon^\mu_{-\kappa}(k_-)
\,\equiv\, \epsilon_{\kappa}^\mu,
\eq
where $\kappa=\pm\,1$ stands for the polarization. The perturbative series for the amplitudes read
\bq
{\cal A}_{\ssP q_\mp}\! =
4\pi\alpha(\mu^2_{P})\,\sum_{n=0}^{\infty}
\bigg(\frac{\alpha}{4\pi}\bigg)^{\!\!n}\! {\cal A}_{\ssP q_\mp}^{(n)}
~~~
(P= T,L),
\qquad
{\cal B}_{\ssT q_-}\! =
4\pi\alpha\,\sum_{n=0}^{\infty}
\bigg(\frac{\alpha}{4\pi}\bigg)^{\!\!n}\! {\cal B}_{\ssT q_-}^{(n)}\,,
\eq
where $\alpha(\mu^2)= {\al_e(\mu^2)}/{\sw^2(\mu^2)}$ 
and  the coupling constants are supposed to be normalized at $\mu=M_W$ 
unless the normalization point is  indicated explicitly. The Born amplitudes read:
\bq
{\cal A}_{\ssT u_-}^{(0)}= \frac{1}{\hat{u}},
\qquad\qquad
{\cal A}_{\ssT d_-}^{(0)}= \frac{1}{\hat{t}},
\qquad\qquad
{\cal B}_{\ssT q_-}^{(0)} = 0,
\nn
\eq
\bq
{\cal A}_{\ssL u_-}^{(0)}\!\!\! =\!
\frac{1}{2\,\hat{s}}\bigg[1+\frac{\tw^2(\mu^2_L)}{3}\bigg],
\quad
{\cal A}_{\ssL d_-}^{(0)}\!\!\! =\!
\frac{1}{2\,\hat{s}}\bigg[\!-1+\frac{\tw^2(\mu^2_L)}{3}\bigg],
\quad
{\cal A}_{\ssL u_+}^{(0)}\!\! =\!
\frac{2}{3\,\hat{s}}\tw^2(\mu^2_L),
\quad
{\cal A}_{\ssL d_+}^{(0)}\!\! =\!
-\,\frac{1}{3\,\hat{s}}\tw^2(\mu^2_L).
\label{born}
\eq
The one-loop contribution to the second transverse Lorentz-Dirac structure of \eqn{dirac-structure} is particularly simple and does not contain Sudakov logarithms
\bq
{\cal B}_{\ssT u_-}^{(1)} =
\bigg[
  \beta_{\ssT,\ssSU}^{(1)}
+ \tw^2\beta_{\ssT,\ssY}^{(1)}
\bigg]\frac{1}{\hat{t}\,\hat{u}},
\qquad\qquad
{\cal B}_{\ssT d_-}^{(1)} =
- \,{\cal B}_{\ssT u_-}^{(1)}(\hat{t}\leftrightarrow\hat{u}),
\eq
\bqa
\beta_{\ssT\!,\ssSU}^{\,(1)} &=&
- \bigg(
    4
  - {5\over2}\frac{\hat{u}}{\hat{t}}
  \bigg)\frac{\hat{u}}{\hat{t}} \big[ L_{us}^2 + \pi^2 \big]
+ 3\frac{\hat{t}}{\hat{u}}\big[  L_{ts}^2 + \pi^2 \big]
- \bigg(
    {9\over2}
  - 5\frac{\hat{u}}{\hat{t}}
  \bigg) L_{us}
+ {5\over2},
\nn\\
\beta_{\ssT,\ssY}^{\,(1)} &=&
- {1\over9}\bigg(
    1
  + {1\over2}\frac{\hat{u}}{\hat{t}}
  \bigg)\frac{\hat{u}}{\hat{t}} \big[ L_{us}^2 + \pi^2 \big]
- {1\over9}\bigg(
    {3\over2}
  + \frac{\hat{u}}{\hat{t}}
  \bigg) L_{us}
- {1\over18}.
\eqa
All the notations  are explained at the end of the section.
The one-loop corrections to the remaining Lorentz-Dirac structure can be
formally decomposed according to the gauge coupling constant factor
\bq
{\cal A}_{\ssP q_\mp}^{(1)} =
  {\cal A}_{\ssP q_\mp\!,\,\ssSU}^{(1)}
+ \tw^2\,{\cal A}_{\ssP q_\mp\!,\ssY}^{(1)}
+ \sw^2\Big(
    {\cal A}_{\ssP q_\mp\!,\,\ssQED}^{(1)}
  - {\cal A}_{\ssP q_\mp\!,\,sub}^{(1)}
  \Big),
\qquad
(P= T,L),
\eq
where the last term ${\cal A}_{\ssP q_\mp\!,\,sub}^{(1)}$ corresponds to
the first order term of the expansion of the singular QED
factor~(\ref{QED})
and  cancels the
infrared logarithms coming from soft photons and from photons collinear to the incoming
quarks\footnote{In contrast to \sect{sec4.1}, we normalize here
${\cal U}(\al_e)\big|_{Q^2=\lm^2=M_W^2}=1$. In this case the
amplitude is manifestly $\lambda$ independent.
The numerical estimates are obtained with the normalization of \sect{sec4.1},
where the  $\lambda$ dependence survives in the imaginary part of the
one-loop amplitude, but does not contribute to the cross section up to N$^3$LL
approximation.}.
The $QED$ correction factorizes with respect to the Born amplitude and reads
\bqa
{\cal A}_{\ssP q_\mp,\ssQED}^{(1)} &=&
\bigg[
  2 \Big( Q_q^2 \!+\! 1 \Big) L_\gamma L
+ 4 Q_q L_\gamma L_{ut}
- Q_q^2 L_\gamma^2
- 2\,Q_q^2 L_\gamma \Lz
- \Big( 3 Q_q^2 \!+\! 2 \Big) L_\gamma
+ \Delta_{\ssP\!,\lambda}^{(1)}
\bigg]{\cal A}_{\ssP q_\mp}^{(0)},
\nn\\
\Delta_{\ssT,\lambda}^{(1)} &=&
  \bigg(
    {9\over2} \!
  - \!{7\over2}\wz \!
  + \!{2\over3}\wz^2
  \bigg) \wz \Lz\!
- \!\bigg[
    \frac{4}{\betaz}
  - \bigg(
      {3\over2}\!
    - \!{13\over6}\wz\!
    + \!{2\over3}\wz^2
    \bigg) \wz\betaz
  \bigg] L_{\chi_{\!_Z}}\!
- \!L_{\chi_{\!_Z}}^2\!
- \!\pi^2\!
+ \!5\wz\!
- \!{4\over3}\wz^2,
\nn\\
\Delta_{\ssL,\lambda}^{(1)} &=&
  \bigg(
    {9\over2}\!
  - \!11\wz \!
  + \!3\wz^2
  \bigg) \wz \Lz\!
- \!\bigg[
    \frac{4}{\betaz}
  + \bigg(
      {1\over2}\!
    + \!5\wz\!
    - \!3\wz^2
    \bigg) \wz\betaz
  \bigg] L_{\chi_{\!_Z}}\!
+ \!L_{\chi_{\!_Z}}^2\!
+ \!\pi^2\!
+ \!11\wz\!
- 6\wz^2,
\nn\\
{\cal A}_{\ssP q_\mp,sub}^{(1)} &=&
\bigg[
- \Big( Q_q^2 \!+\! 1 \Big) \ln^2\frac{Q^2}{\lambda^2}
+ \Big( 3 Q_q^2 \!-\! 4 Q_q L_{ut} \Big) \ln\frac{Q^2}{\lambda^2}
+ \ln^2\frac{M_W^2}{\lambda^2}
+ 2 \ln\frac{M_W^2}{\lambda^2}
\bigg]{\cal A}_{\ssP q_\mp}^{(0)}.
\label{qed}
\eqa
where  $Q^2= -\hat{s}-i\,0^+$ and  $Q_q$ is the electric charge of the
quark. Note that ${\cal A}_{\ssP q_\mp,\ssQED}^{(1)}$ vanishes for
$M_Z\sim\lambda\to 0$.
After the subtraction the $\lambda$-dependence disappears
and we get the QED contribution in terms of the parameters of the evolution equation
\bq
{\cal A}_{\ssP q_\mp,\ssQED}^{(1)}- {\cal A}_{\ssP q_\mp,sub}^{(1)}=
-\bigg[
  { \gamma_{q,\ssQED}^{(1)} \over 2}\, L^2
+ \Big( \zeta_{q,\ssQED}^{(1)} + \xi_{q,\ssQED}^{(1)} + \chi_{_{q,\ssQED}}^{(1)}
  \Big) L
+ \Delta_{\ssP q,\ssQED}^{(1)}
\bigg] {\cal A}_{\ssP q_\mp}^{(0)},
\nn
\eq
\bq
\gamma_{q,\ssQED}^{(1)} =
- 2 \Big( Q_q^2 \!+\! 1 \Big),
\qquad\;
\zeta_{q,\ssQED}^{(1)} =
3 Q_q^2,
\qquad\;
\chi_{_{q,\ssQED}}^{(1)} =
- 4 Q_q L_{ut},
\qquad\;
\xi_{q,\ssQED}^{(1)}\! =
2 \Big( Q_q^2 \!+\! 1 \Big) \Lz,
\nn
\eq
\bq
\Delta_{\ssP q,\ssQED}^{(1)} =
- \Big( 3 Q_q^2 \!+\! 2 \Big) \Lz
+ 4 Q_q \Lz L_{ut}
- Q_q^2 \Lz^2
- \Delta_{\ssP\!,\lambda}^{(1)}.
\eq
In order to present the result of the remaining $SU(2)$ and $Y$ components,
in terms of the coefficients of the evolution equations, it is necessary to
analyze the isospin structure of the amplitude for left handed quarks.
The general $SU(2)$ basis for the amplitude $A_T$ of the left-handed
quark-antiquark pair transition into two transverse gauge bosons reads
\bq
\bar{q}_\ssL\,q_\ssL \to B_a\,B_b:\qquad
(\bar{u}_+ \;\, \bar{d}_+)\,\bigg(
  {\cal A}_{1}\,\frac{\bfm{\sigma}_a}{2}\frac{\bfm{\sigma}_b}{2}
+ {\cal A}_{2}\,\frac{\bfm{\sigma}_b}{2}\frac{\bfm{\sigma}_a}{2}
+ {\cal A}_{3}\,\delta_{ab}\I
\bigg)\,
\left(\ba{c}u_-\\d_-\ea\right),
\qquad
a,b= 1,2,3,
\label{colorT}
\eq
where
$B_a$ are $SU(2)$ gauge fields and $\sigma_a$ are the Pauli matrices.
From the definition $W_T^\pm=(B_1\mp i B_2)/\sqrt{2}$ we get the following
structure for the production of $W_T^+ W_T^-$
\bqa
\bar{q}_\ssL\,q_\ssL \to W_T^+W_T^-:\qquad&&
(\bar{u}_+ \;\, \bar{d}_+)\,\bigg(
  {\cal A}_{1}\,\frac{\bfm{\sigma}_-}{2}\frac{\bfm{\sigma}_+}{2}
+ {\cal A}_{2}\,\frac{\bfm{\sigma}_+}{2}\frac{\bfm{\sigma}_-}{2}
+ {\cal A}_{3}\,\I
\bigg)\,
\left(\ba{c}u_-\\d_-\ea\right) =
\nn\\
&&
(\bar{u}_+ \;\, \bar{d}_+)\,\bigg[
  \frac{{\cal A}_{1}}2\left(\ba{cc}0&0\\0&1\ea\right)
+ \frac{{\cal A}_{2}}2\left(\ba{cc}1&0\\0&0\ea\right)
+ {\cal A}_{3}\left(\ba{cc}1&0\\0&1\ea\right)
\bigg]\,
\left(\ba{c}u_-\\d_-\ea\right),
\label{colorWT}
\eqa
where $\bfm{\sigma}_\pm=(\bfm{\sigma}_1\pm i\,\bfm{\sigma}_2)/\sqrt{2}$.
Thus
\bq
{\cal A}_{\ssT u_-}= \frac{1}{2}{\cal A}_{2} + {\cal A}_{3},
\qquad
{\cal A}_{\ssT d_-}= \frac{1}{2}{\cal A}_{1} + {\cal A}_{3}.
\label{AT}
\eq
We introduce now the isospin vector amplitude $\bfm{\cal A}_\ssT$ of
Section~\ref{sec3}
\bq
\bfm{\cal A}_{\ssT q_-} =
\left(\!\ba{c}
{\cal A}_{1}
\\
{\cal A}_{2}
\\
{\cal A}_{3}
\ea\!\right).
\eq
The Born amplitudes ${\cal A}_{\ssT u_-}^{(0)}$ and ${\cal A}_{\ssT d_-}^{(0)}$
of \eqn{born} correspond to the vector
\bq
\bfm{\cal A}_{\ssT q_-}^{(0)} =
2
\left(\ba{c}
\displaystyle{ 1/\hat{t} }
\\
\displaystyle{ 1/\hat{u} }
\\
\displaystyle{ 0 }
\ea\right).
\eq
For the amplitude $A_{\ssL q_-}$ of the left-handed quark-antiquark
pair transition into Goldstone bosons, the isospin basis is:
\bq
  {\cal A}_\sigma\,
  (\bar{u}_+ \;\, \bar{d}_+)\frac{\bfm{\sigma}_a}{2}\!
  \left(\!\ba{c}u_-\\d_-\ea\!\right)
  (\Phi_0 \;\, \Phi_{_+})\frac{\bfm{\sigma}_a}{2}\!
  \left(\!\ba{c}\Phi_0^*\\\Phi_{_-}\ea\!\right)
+ {\cal A}_\mathds{1}\,
  (\bar{u}_+ \;\, \bar{d}_+)\I\!
  \left(\!\ba{c}u_-\\d_-\ea\!\right)
  (\Phi_0 \;\, \Phi_{_+})\I\!
  \left(\!\!\ba{c}\Phi_0^*\\\Phi_{_-}\ea\!\!\right).
\label{colorL}
\eq
In the first term the sum over $a$ goes from $1$ to $3$, but only $\sigma_3$
contributes to the production of charged $\phi$ pair so that
\bq
{\cal A}_{\ssL u_-}= - \frac{1}{4}\,{\cal A}_\sigma + {\cal A}_\mathds{1},
\qquad
{\cal A}_{\ssL d_-}= \frac{1}{4}\,{\cal A}_\sigma + {\cal A}_\mathds{1}.
\label{AL}
\eq
As for the transverse case we introduce the isospin vector amplitude
$\bfm{\cal A}_\ssL$ of Section~\ref{sec3}
\bq
\bfm{\cal A}_{\ssL q_-} =
\left(\!\ba{c}
{\cal A}_\sigma
\\
{\cal A}_\mathds{1}
\ea\!\right).
\eq
The Born amplitudes ${\cal A}_{\!\ssL u_-}^{(0)}$ and ${\cal A}_{\!\ssL d_-}^{(0)}$
of \eqn{born} correspond to the vector
\bq
\bfm{\cal A}_{\ssL q_-}^{(0)}\!\! =
-\frac{2}{\hat{s}}\!
\left(\ba{c}
\displaystyle{ 1 }
\\
\displaystyle{ \frac{Y_iY_f}{4}\tw^2(\mu_L^2) }
\ea\right) =
\frac{2}{\hat{s}}\!
\left(\ba{c}
\displaystyle{ -\,1 }
\\
\displaystyle{ \frac{1}{12}\tw^2(\mu_L^2) }
\ea\right),
\eq
where $Y_i$ ($Y_f$) is the hypercharge of the initial (final) state.
The isospin vector amplitudes can also be decomposed according to the gauge
couplings
\bq
\bfm{\cal A}_{\ssP q_-}^{(1)} =
  \bfm{{\cal A}}_{\ssP q_-,\ssSU}^{(1)}
+ \tw^2\,\bfm{{\cal A}}_{\ssP q_-,\ssY}^{(1)}
+ \sw^2\Big(
    \bfm{{\cal A}}_{\ssP q_-,\,\ssQED}^{(1)}
  - \bfm{{\cal A}}_{\ssP q_-,\,sub}^{(1)}
  \Big),
\qquad (P= T,L).
\eq
The expressions for the $SU(2)$ and $Y$ components of the vectorial amplitude
are then written in terms of the parameters of the evolution equation
\bq
\bfm{{\cal A}}_{\ssP q_-,i}^{(1)} =
  \bigg[
    { \gamma_{\ssP q_-,i}^{(1)} \over 2}\, L^2
  + \Big(
      \zeta_{\ssP q_-,i}^{(1)}
    + \xi_{\ssP q_-,i}^{(1)}
    + \bfm{\chi}_{_{\ssP q_-,i}}^{(1)}
    \Big) L
  + \bfm{\Delta}_{\ssP q_-,i}^{(1)},
  \bigg]  \bfm{{\cal A}}_{\ssP q_-,i}^{(0)},
\qquad
(i=SU(2),Y).
\label{ad}
\eq
The coefficients $\gamma_{\ssP,i}^{(1)}$, $\zeta_{\ssP,i}^{(1)}$
and $\bfm{\chi}_{_{\ssP,i}}^{(1)}$ are universal, the other are obtained
by explicit calculation.

The expression for the  amplitudes ${\cal A}_{\ssT q_-,i}^{(1)}$ and
${\cal A}_{\ssL q_-,i}^{(1)}$ in terms of the anomalous dimensions
can be then obtained by using Eqs.(\ref{AT},~\ref{AL}) and takes the form:
\bq
{\cal A}_{\ssP q_-,i}^{(1)}\! =\!
  \bigg[
    { \gamma_{\ssP q_-,i}^{(1)} \over 2} L^2
  + \bigg(
      \zeta_{\ssP q_-,i}^{(1)}
    + \xi_{\ssP q_-,i}^{(1)}
    + \chi_{_{\ssP q_-,i}}^{(1)}
    \bigg)\! L
  + \Delta_{\ssP q_-,i}^{(1)}
  \bigg] {\cal A}_{\ssP q_-}^{(0)}
  + \bigg[
    \bar{\chi}_{_{\ssP q_-,i}}^{(1)} L
  + \bar{\Delta}_{\ssP q_-,i}^{(1)}
  \bigg] \bar{\cal A}_{\ssP q_-}^{(0)},\quad
\eq
where we introduced the notations
\bq
\bar{\cal A}_{\ssP u_\mp}^{(0)} =
{\cal A}_{\ssP d_\mp}^{(0)},
\qquad\qquad
\bar{\cal A}_{\ssP d_\mp}^{(0)} =
{\cal A}_{\ssP u_\mp}^{(0)}.
\eq
The coefficient are given by:

\vspace{.5cm}
\noindent
\underline{Transverse $W$, left-handed quarks, SU(2) component:}
\bqa
\gamma_{\ssT q_-,\ssSU}^{(1)}\! &=&
- {11\over2},
\qquad\;
\zeta_{\ssT q_-,\ssSU}^{(1)}\! =
\frac{9}{4},
\qquad\;
\xi_{\ssT q_-,\ssSU}^{(1)}\! =
{5\over2} \Lz,
\qquad\;
\bfm{\chi}_{_{\ssT q_-,\ssSU}}^{(1)}\! =
\!\left(\!
\ba{ccc}
\displaystyle{
- 4 L_{ts}
}
&
\displaystyle{
0
}
&
\displaystyle{
4 L_{ut}
}
\\
\displaystyle{
0
}
&
\displaystyle{
- 4 L_{us}\;
}
&
\displaystyle{
4 L_{tu}
}
\\
\displaystyle{
L_{us}
}
&
\displaystyle{
L_{ts}
}
&
\displaystyle{
0
}
\ea
\!\right)\!,
\nn\\[+.3cm]
\chi_{_{\ssT u_-,\ssSU}}^{(1)}\!\! &=&
2 L_{ts} - 4 L_{us},
\qquad
\chi_{_{\ssT d_-,\ssSU}}^{(1)}\!\! =
2 L_{us} - 4 L_{ts},
\qquad
\bar{\chi}_{_{\ssT u_-,\ssSU}}^{(1)}\!\! =
2 L_{us},
\qquad
\bar{\chi}_{_{\ssT d_-,\ssSU}}^{(1)}\!\! =
2 L_{ts},
\nn\\[+.2cm]
&&
\Delta_{\ssT u_-,\ssSU}^{(1)} =
\Delta_{\ssT,\ssSU}^{(1)},
\qquad\qquad
\Delta_{\ssT d_-,\ssSU}^{(1)} =
\Delta_{\,\ssT\!,\ssSU}^{(1)}(\hat{t}\leftrightarrow\hat{u}),
\qquad\qquad
\bar{\Delta}_{\ssT q_-,\ssSU}^{(1)} =
0,
\nn\\[+.2cm]
{\Delta}_{\,\ssT\!,\ssSU}^{(1)} &=&
- {5\over4}\bigg(
    1\!
  - \!{3\over5}\frac{\hat{u}}{\hat{t}}\!
  + \!\frac{\hat{u}^2}{\hat{t}^2}
  \bigg) L_{us}^2
+ {3\over2} L_{ts}^2
+ 2\bigg(
    1\!
  + \!\frac{\hat{u}}{\hat{t}}
  \bigg) L_{ts} L_{us}
+ \bigg(
    {9\over4}\!
  - \!{5\over2}\frac{\hat{u}}{\hat{t}}\!
  \bigg)\! L_{us}
+ 2 \Lz L_{ut}
+ {485\over72}
\nn\\
&&
+ {\pi^2\over4}\bigg(
    7
  + 3\frac{\hat{u}}{\hat{t}}
  - 5\frac{\hat{u}^2}{\hat{t}^2}
  \bigg)
- {11\over2}\wz
+ {3\over2}\wz^2
- {1\over2}\wh
+ {1\over6}\wh^2
- {1\over2}\wt
- \wt^2
+ \Lt
+ \Big(\!
    1\!
  - \!\wt^3
  \!\Big) \Ltw
\nn\\
&&
- \,{1\over4}\bigg(
    {1\over3}\!
  + \!17\wz\!
  - \!{31\over2}\wz^2\!
  + \!3\wz^3
  \bigg) \Lz
- {1\over4} \Lz^2
+ \,{1\over2}\bigg(\!
    1\!
  - \!{3\over2}\wh\!
  + \!{3\over4}\wh^2\!
  - \!{1\over6}\wh^3
  \bigg) \Lh\!
+ L_{\chi_{\!_Z}}^2
\nn\\
&&
+ \bigg[
    {9\over2}\frac1{\betaz}
  - \bigg(
      1\!
    - \!{19\over8}\wz\!
    + \!{3\over4}\wz^2
    \bigg)\wz\betaz
  \bigg]\! L_{\chi_{\!_Z}}
- \!{1\over2}\bigg[
    \frac1{\betah}\!
  + \!\bigg(\!
      1\!
    - \!{5\over12}\wh\!
    + \!{1\over6}\wh^2
    \bigg)\wh\betah
  \bigg]\! L_{\chi_{\!_H}}.
\label{trsu2}
\eqa

\vspace{.3cm}
\noindent
\underline{Transverse $W$, left-handed quarks, Y component:}
\bqa
\gamma_{\ssT q_-,\ssY}^{(1)}\! &=&
- {1\over18},
\qquad\;
\zeta_{\ssT q_-,\ssY}^{(1)}\! =
\frac{1}{12},
\qquad\;
\xi_{\ssT q_-,\ssY}^{(1)}\! =
{1\over18} \Lz,
\qquad\;
\bfm{\chi}_{_{\ssT q_-,\ssY}}^{(1)}\! = 0,
\qquad\;
\chi_{_{\ssT q_-,\ssY}}^{(1)}\! =
\bar{\chi}_{_{\ssT q_-,\ssY}}^{(1)}\! =
0,
\nn\\[+.2cm]
&&
\Delta_{\ssT u_-,\ssY}^{(1)} =
\Delta_{\ssT,\ssY}^{(1)},
\qquad\qquad\qquad
\Delta_{\ssT d_-,\ssY}^{(1)} =
\Delta_{\ssT,\ssY}^{(1)}(\hat{t}\leftrightarrow\hat{u}),
\qquad\qquad\qquad
\bar{\Delta}_{\ssT q_-,\ssY}^{(1)} =
0,
\nn\\[+.1cm]
{\Delta}_{\ssT,\ssY}^{(1)} &=&
  {1\over36}\bigg(
    1
  + 3\frac{\hat{u}}{\hat{t}}
  + \frac{\hat{u}^2}{\hat{t}^2}
  \bigg) L_{us}^2
+ {1\over12}\bigg(
    1
  + {2\over3}\frac{\hat{u}}{\hat{t}}
  \bigg) L_{us}
+ {67\over72}
+ {\pi^2\over108}\bigg(
    1
  + 9\frac{\hat{u}}{\hat{t}}
  + 3\frac{\hat{u}^2}{\hat{t}^2}
  \bigg)
\nn\\
&&
- {1\over36} \Lz^2
+ {1\over2}\bigg(
    {5\over6}
  - \wz
  \bigg) \Lz
- {1\over2}\bigg(
    \frac1{\betaz}
  + \wz\betaz
  \bigg) L_{\chi_{\!_Z}}.
\eqa

\vspace{.3cm}
\noindent
\underline{Longitudinal $W$, left-handed quarks, SU(2) component:}
\vspace{-.3cm}
\bqa
\gamma_{\ssL q_-,\ssSU}^{(1)}\! &=&\!
- \,3,
\quad\;\;\;
\zeta_{\ssL q_-,\ssSU}^{(1)}\! =\!
\frac{21}{4} - {3\over2}\wt,
\quad\;\;\;
\xi_{\ssL q_-,\ssSU}^{(1)}\! =\! \Lz,
\quad\;\;\;
\bfm{\chi}_{_{\ssL q_-,\ssSU}}^{(1)}\!\! =\!\!
\left(\!
\ba{cc}
\displaystyle{
- 2 L_{us} \!-\! 2 L_{ts}
}
\;\;
&
\displaystyle{
4 L_{ut}
}
\\[+0.2cm]
\displaystyle{
\frac{3}{4}\,L_{ut}
}
&
\displaystyle{
0
}
\ea
\!\right)\!\!,
\!\!\!
\nn\\[+.1cm]
\chi_{_{\ssL u_-,\ssSU}}^{(1)}\!\! &=&
L_{ts} - 3 L_{us},
\qquad\;
\chi_{_{\ssL d_-,\ssSU}}^{(1)}\!\! =
L_{us} - 3 L_{ts},
\qquad\;
\bar{\chi}_{_{\ssL u_-,\ssSU}}^{(1)}\!\! =
2 L_{us},
\qquad\;
\bar{\chi}_{_{\ssL d_-,\ssSU}}^{(1)}\!\! =
2 L_{ts},
\nn\\[+.2cm]
\bfm{\Delta}_{\ssL q_-,\ssSU}^{(1)}\! &=&
\left(
\ba{cc}
\displaystyle{
- {1\over2}L_+\!
+ \!{575\over72}\!
+ \!{5\over3}\pi^2\!
+ \!{19\over6}i\pi
}
\qquad
&
\displaystyle{
  L_-\!
- \!4 L_{ut} \Lz\!
- \!{416\over3}\!
- \!82i\pi
}
\\[+0.3cm]
\displaystyle{
  {3\over16} L_-
- {1\over4} L_{ut} \Lz
}
&
\displaystyle{
- {43\over24}
- {1\over3}\pi^2
}
\ea
\right)
+ \Delta_{\ssL,\ssSU}^{\!(1)}\I,
\nn\\[+.2cm]
\Delta_{\ssL u_-\!,\ssSU}^{(1)}\! &=&
  \;\;{3\over4}\,\frac{\hat{s}}{\hat{t}}\, \Big( L_{us}^2 \!+\! \pi^2 \Big)
- {1\over4}\,\frac{\hat{s}}{\hat{u}}\, \Big( L_{ts}^2 \!+\! \pi^2 \Big)
- L_{tu}\, \Lz
+ {1471\over72}
+ {2\over3}\,\pi^2
+ {71\over6}\,i\pi
+ \Delta_{\ssL,\ssSU}^{(1)},
\nn\\[+.2cm]
\Delta_{\ssL d_-\!,\ssSU}^{(1)}\! &=&
  \;\;{3\over4}\,\frac{\hat{s}}{\hat{u}}\, \Big( L_{ts}^2 \!+\! \pi^2 \Big)
- {1\over4}\,\frac{\hat{s}}{\hat{t}}\, \Big( L_{us}^2 \!+\! \pi^2 \Big)
- L_{ut}\, \Lz
- {1025\over72}
+ {2\over3}\,\pi^2
- {26\over3}\,i\pi
+ \Delta_{\ssL,\ssSU}^{(1)},
\nn\\[+.2cm]
\bar{\Delta}_{\!\ssL u_-\!,\ssSU}^{(1)}\! &=&\!
- {1\over2}\frac{\hat{s}}{\hat{t}} \Big(\! L_{us}^2 \!+\! \pi^2 \!\Big)\!
+ \!{112\over9}\!
- \!\pi^2\!
+ \!{26\over3}i\pi,
\qquad\;
\bar{\Delta}_{\!\ssL d_-\!,\ssSU}^{(1)}\! =\!
- {1\over2}\frac{\hat{s}}{\hat{u}} \Big(\! L_{ts}^2 \!+\! \pi^2 \!\Big)\!
- \!{200\over9}\!
- \!\pi^2\!
- \!{71\over6}i\pi,\!\!\!\!\!
\nn\\[+.2cm]
\Delta_{\ssL,\ssSU}^{(1)} &=&
- 17\wz
+ {27\over4}\wz^2
- \wh
+ {1\over4}\wh^2
+ {3\over2}\wt
- {3\over2}\wt^2
- {1\over4}\bigg(
    13\!
  + \!39\wz\!
  - \!{117\over2}\wz^2\!
  + \!{27\over2}\wz^3
  \bigg) \Lz
\nn\\
&&
+ {1\over2}\bigg(
    1\!
  - \!3\wh\!
  + \!{5\over4}\wh^2\!
  - \!{1\over4}\wh^3
  \bigg) \Lh
+ {3\over2}\wt \Lt
+ {3\over2}\Big(\!
    1\!
  - \!\wt^2
  \!\Big)\wt \Ltw
- {1\over4} \Lz^2
+ {1\over4} L_{\chi_{\!_Z}}^2
+ {1\over2} L_{\chi_{\!_H}}^2
\nn\\
&&
+ \bigg[
    {9\over2}\frac1{\betaz}\!
  + \!\bigg(
      {1\over4}\!
    + \!{63\over8}\wz\!
    - \!{27\over8}\wz^2
    \bigg)\wz\betaz
  \bigg] L_{\chi_{\!_Z}}\!
- \!\bigg[
    {1\over2}\frac1{\betah}\!
  + \!\bigg(
      1\!
    - \!{3\over8}\wh\!
    + \!{1\over8}\wh^2
    \bigg)\wh\betah
  \bigg] L_{\chi_{\!_H}}.
\eqa

\vspace{.3cm}
\noindent
\underline{Longitudinal $W$, left-handed quarks, Y component:}
\bqa
\gamma_{\ssL q_-,\ssY}^{(1)} &=&
- \frac{5}{9},
\qquad\qquad
\zeta_{\ssL q_-,\ssY}^{(1)} =
\frac{13}{12},
\qquad\qquad
\xi_{\ssL q_-,\ssY}^{(1)}\! =
{5\over9} \Lz,
\qquad\qquad
\bfm{\chi}_{_{\ssL q_-,\ssY}}^{(1)} =
- \frac{1}{3} L_{ut}\,\I,
\nn\\[+.1cm]
\chi_{_{\ssL\!q_-,\ssY}}^{(1)} &=&
- {1\over3} L_{ut},
\qquad
\bar{\chi}_{_{\ssL\!q_-,\ssY}}^{(1)} =
0,
\qquad\;
\bfm{\Delta}_{\ssL q_-,\ssY}^{(1)} =
\Delta_{\ssL,\ssY}^{(1)}\,\I,
\qquad\;
\Delta_{\ssL q_-\!,\ssY}^{(1)}\! =
\Delta_{\ssL\!,\ssY}^{(1)},
\qquad\;
\bar{\Delta}_{\ssL q_-\!,\ssY}^{(1)}\! =
0,
\nn\\[+.1cm]
\Delta_{\ssL,\ssY}^{(1)} &=&
- {1\over12} L_-
+ {1\over3}  L_{ut} \Lz
- {731\over72}
- {5\over27}\pi^2
- {41\over6}i\pi
+ {1\over2}\bigg(
    {5\over6}
  - {7\over2}\wz
  + {1\over2}\wz^2
  \bigg) \Lz
\nn\\
&&
- {1\over36} \Lz^2
- {1\over4} L_{\chi_{\!_Z}}^2
- {1\over4}\bigg[
    2\frac1{\betaz}
  + \Big(\!
      5\!
    - \!\wz
    \!\Big)\wz\betaz
  \bigg] L_{\chi_{\!_Z}}
- {1\over2}\wz.
\eqa
For the right-handed quarks the amplitudes are isospin singlet so that
the $SU(2)$ and $Y$ contributions factorize with respect to the
Born amplitude:
\bq
{\cal A}_{\ssL q_+,i}^{(1)} =
\bigg[
  { \gamma_{\ssL q_+,i}^{(1)} \over 2}\, L^2
+ \Big(
    \zeta_{\ssL q_+,i}^{(1)}
  + \xi_{\ssL q_+,i}^{(1)}
  + \chi_{_{\ssL q_+,i}}^{(1)}
  \Big) L
+ \Delta_{\ssL q_+,i}^{(1)}
\bigg]{\cal A}_{\ssL q_+}^{(0)},
\qquad\quad
i= SU(2),Y.
\label{adR}
\eq

\vspace{.3cm}
\noindent
\underline{Longitudinal $W$, right-handed quarks, SU(2) component:}
\bqa
\gamma_{\ssL q_+,\ssSU}^{(1)} &=&
- {3\over2},
\qquad\quad
\zeta_{\ssL q_+,\ssSU}^{(1)} =
3 - {3\over2}\wt,
\qquad\quad
\chi_{_{\ssL q_+,\ssSU}}^{(1)} =
0,
\qquad\quad
\xi_{\ssL q_+,\ssSU}^{(1)} =
{1\over2} \Lz,
\nn\\[+.2cm]
\Delta_{\ssL q_+,\ssSU}^{(1)} &=&
- 17\wz
+ {27\over4}\wz^2
- \wh
+ {1\over4}\wh^2
+ {3\over2}\wt
- {3\over2}\wt^2
+ {5\over6}
+ {1\over6} \pi^2
+ {3\over2}\Big(\!
    1\!
  - \!\wt^2
  \!\Big)\wt \Ltw
+ {3\over2}\wt \Lt
\nn\\
&&
- {1\over4}\bigg(
    10\!
  + \!39\wz\!
  - \!{117\over2}\wz^2\!
  + \!{27\over2}\wz^3
  \bigg) \Lz
+ {1\over2}\bigg(
    1\!
  - \!3\wh\!
  + \!{5\over4}\wh^2\!
  - \!{1\over4}\wh^3
  \bigg) \Lh
+ {1\over4} L_{\chi_{\!_Z}}^2
+ {1\over2} L_{\chi_{\!_H}}^2
\nn\\
&&
+ \bigg[
    {9\over2}\frac1{\betaz}\!
  + \!\bigg(
      {1\over4}\!
    + \!{63\over8}\wz\!
    - \!{27\over8}\wz^2
    \bigg)\wz\betaz
  \bigg] L_{\chi_{\!_Z}}\!
- \!\bigg[
    {1\over2}\frac1{\betah}\!
  + \!\bigg(
      1\!
    - \!{3\over8}\wh\!
    + \!{1\over8}\wh^2
    \bigg)\wh\betah
  \bigg] L_{\chi_{\!_H}}.
\eqa

\vspace{.3cm}
\noindent
\underline{Longitudinal $W$, right-handed quarks, Y component:}
\bqa
\gamma_{\ssL u_+,\ssY}^{(1)} &=&
- \frac{25}{18},
\qquad\qquad
\zeta_{\ssL u_+,\ssY}^{(1)} =
\frac{7}{3},
\qquad\qquad
\chi_{_{\ssL u_+,\ssY}}^{(1)} =
- \frac{4}{3} L_{ut},
\qquad\qquad
\xi_{\ssL u_+,\ssY}^{(1)}\! =
{25\over18}  \Lz,
\nn\\[+.2cm]
\Delta_{\ssL u_+,\ssY}^{(1)} &=&
- {1\over3} L_-
+ {4\over3}  L_{ut} \Lz
- {209\over18}
- {25\over54}\pi^2
- {1\over2}\wz
- \bigg(
    {5\over6}\!
  - \!{7\over4}\wz\!
  + \!{1\over4}\wz^2
  \bigg) \Lz
- {4\over9} \Lz^2
\nn\\
&&
- {1\over4} L_{\chi_{\!_Z}}^2\!
- \!{1\over4}\bigg[
    \frac{2}{\betaz}
  + \Big(\!
      5\!
    - \!\wz
    \!\Big)\wz\betaz
  \bigg] L_{\chi_{\!_Z}},
\nn\\[+.2cm]
\gamma_{\ssL d_+,\ssY}^{(1)} &=&
- \frac{13}{18},
\qquad\qquad
\zeta_{\ssL d_+,\ssY}^{(1)} =
\frac{4}{3},
\qquad\qquad
\chi_{_{\ssL d_+,\ssY}}^{(1)} =
\frac{2}{3} L_{ut},
\qquad\qquad
\xi_{u_+,\ssL,\ssY}^{(1)}\! =
{13\over18} \Lz,
\nn\\[+.2cm]
\Delta_{\ssL d_+,\ssY}^{(1)} &=&
  {1\over6} L_-
- {2\over3}  L_{ut} \Lz
- {94\over9}
+ {13\over54}\pi^2
- {41\over6}i\pi
- {1\over2}\wz
+ \bigg(
    {1\over6}
  - {7\over4}\wz
  + {1\over4}\wz^2
  \bigg) \Lz
- {1\over9} \Lz^2
\nn\\
&&
- {1\over4} L_{\chi_{\!_Z}}^2
- {1\over4}\bigg[
    \frac{2}{\betaz}
  + \Big(\!
      5\!
    - \!\wz
    \!\Big)\wz\betaz
  \bigg] L_{\chi_{\!_Z}}.
\eqa

\vspace{.5cm}
The NNLL  two-loop amplitudes are obtained by using
Eqs.(\ref{expsola},\ref{AT},\ref{AL}):
\bq
{\cal B}_{\ssT q_-}^{(2)} =
  \frac{\gamma_{\ssT q_-}^{(1)}}{2}{\cal B}_{\ssT q_-}^{(1)}\,L^2,
\nn
\eq
\vspace{-.5cm}
\bqa
{\cal A}_{\ssP q_\mp}^{(2)}\!\!
&=&
  \Bigg\{
    {1 \over 8}\Big(\! \gamma_{\ssP q_\mp}^{(1)} \!\Big)^{\!2}\! L^4
  + \bigg[
      \frac{\gamma_{\ssP q_\mp}^{(1)}}{2}
      \Big(\!
        \zeta_{\ssP q_\mp}^{(1)}\!
      + \!\xi_{\ssP q_\mp}^{(1)}\!
      + \!\chi_{\ssP q_\mp}^{(1)}
      \!\Big)\!
    - \!\frac{1}{6}[\beta\gamma]_{\ssP q_\mp}^{(1)}
    \!\bigg] L^3
  + {1 \over 2}\bigg[
      \gamma_{\ssP q_\mp}^{(2)}
    + \Big( \zeta_{\ssP q_\mp}^{(1)}\! + \!\xi_{\ssP q_\mp}^{(1)} \Big)^{\!2}
\nl
&&\quad
    + \,2\,\Big( \zeta_{\ssP q_\mp}^{(1)}\! + \!\xi_{\ssP q_\mp}^{(1)} \Big)
      \chi_{\ssP q_\mp}^{(1)}
    + [\chi^2]_{\ssP q_\mp}^{(1)}
    - [\beta\zeta]_{\ssP q_\mp}^{(1)}
    - [\beta\chi]_{\ssP q_\mp}^{(1)}
    + \Delta_{\ssP q_\mp}^{\!Y\!uk}
    + \gamma_{\ssP q_\mp}^{(1)}\Delta_{\ssP q_\mp}^{(1)}
    \bigg] L^2
  \Bigg\} {\cal A}_{\ssP q_\mp}^{(0)}
\nl
&&\!\!\!\!\!
+ \,\Bigg\{
 \frac{\gamma_{\ssP q_\mp}^{(1)}}{2} \bar{\chi}_{\ssP q_\mp}^{(1)}L^3\!
  + {1 \over 2}\bigg[
      2\Big(\! \zeta_{\ssP q_\mp}^{(1)}\!\! + \!\xi_{\ssP q_\mp}^{(1)} \!\Big)
      \bar{\chi}_{\ssP q_\mp}^{(1)}\!
    + \![\bar{\chi}^2]_{\ssP q_\mp}^{(1)}\!
    - \![\beta\bar{\chi}]_{\ssP q_\mp}^{(1)}\!
    + \!\gamma_{\ssP q_\mp}^{(1)}\bar{\Delta}_{\!\ssP q_\mp}^{(1)}
    \bigg]\! L^2
  \Bigg\} \bar{\cal A}_{\ssP q_\mp}^{(0)}\!
\eqa
where
\bq
f_{\!\ssP q_\mp}^{(1)}\! =
  f_{\!\ssP q_\mp,\ssSU}^{(1)}
+ \tw^2f_{\!\ssP q_\mp,\ssY}^{(1)}
- \sw^2f_{\!q,\ssQED}^{(1)},
\qquad\;
f\!= \!\gamma,\zeta,\xi,\chi,\Delta;
\qquad
\bar{\chi}_{\!\ssP q_-}^{(1)}\! = \bar{\chi}_{\!\ssP q_-,\ssSU}^{(1)},
\qquad
\bar{\chi}_{\!\ssP q_+}^{(1)}\! = 0,
\nn
\eq
\vspace{-.2cm}
\bq
[\beta\! f]_{\!\ssP q_\mp}^{(1)}\!\! = \!
  {19\over6}f_{\!\ssP\!q_\mp,\ssSU}^{(1)}\!
- \!{41\over6}\tw^4 f_{\!\ssP\!q_\mp,\ssY}^{(1)}\!
+ \!{80\over9}\sw^4 f_{\!q,\ssQED}^{(1)},
\quad
f\!\! = \!\gamma,\zeta,\chi;
\quad\;\;
[\beta\bar{\chi}]_{\!\ssP q_-}^{(1)}\!\!\! = \!
{19\over6}\bar{\chi}_{\!\ssP\!q_-,\ssSU}^{(1)},
\quad
[\beta\bar{\chi}]_{\!\ssP q_+}^{(1)}\!\!\! = \!0,
\nn
\eq
\vspace{-.1cm}
\bq
\bar{\Delta}_{\!\ssL q_-}^{(1)}\! = \bar{\Delta}_{\!\ssL q_-,\ssSU}^{(1)},
\qquad\quad
\bar{\Delta}_{\!\ssT q_-}^{(1)} =
\bar{\Delta}_{\!\ssL q_+}^{(1)} = 0,
\qquad\quad
\Delta_{\ssT q_-}^{\!\!Y\!uk} = 0;
\qquad\quad
\Delta_{\ssL q_\pm}^{\!\!Y\!uk} =
- {1\over4} \wt\!
- 3\wt^2,
\eq
with $P = T,L$ and $q=u,d$.
Note that the Yukawa contribution of \eqn{yukawa} is partially contained
already in $\zeta_{L,\ssSU}^{(1)}$ and the remaining piece is given by
$\Delta_{\ssL q_-}^{\!\!Y\!uk}$.
The square of the matrix $\bfm{\chi}$ gives the following
contribution:
\bq
[\chi^2]_{\!\ssP q_\mp}^{(1)}\! =
  [\chi^2]_{\!\ssP q_\mp,\ssSU}^{(1)}
+ 2 \chi_{\!\ssP q_\mp,\ssSU}^{(1)}
  \bigg(
    \tw^2 \chi_{\!\ssP q_\mp,\ssY}^{(1)}
  - \sw^2 \chi_{\!q,\ssQED}^{(1)}
  \bigg)
+ \bigg(
    \tw^2 \chi_{\!\ssP q_\mp,\ssY}^{(1)}
  - \sw^2 \chi_{\!q,\ssQED}^{(1)}
  \bigg)^2
\nn
\eq
\bq
[\chi^2]_{_{\ssT u_-,\ssSU}}^{(1)}\!\! =
16 L_{us}^2 - 12 L_{ts} L_{us} + 4 L_{ts}^2,
\qquad\qquad\qquad
[\chi^2]_{_{\ssT d_-,\ssSU}}^{(1)}\!\! =
4 L_{us}^2 - 12 L_{ts} L_{us} + 16 L_{ts}^2,
\nn
\eq
\vspace{-.1cm}
\bq
[\chi^2]_{_{\ssL u_-,\ssSU}}^{(1)}\!\! = 9 L_{us}^2 - 2 L_{ts} L_{us} + L_{ts}^2,
\qquad
[\chi^2]_{_{\ssL d_-,\ssSU}}^{(1)}\!\! = L_{us}^2 - 2 L_{ts} L_{us} + 9 L_{ts}^2,
\qquad
[\chi^2]_{_{\ssL q_+,\ssSU}}^{(1)}\!\! = 0,
\nn
\eq
\vspace{-.1cm}
\bq
[\bar{\chi}^2]_{\!\ssP q_\mp}^{(1)}\! =
  [\bar{\chi}^2]_{\!\ssP q_\mp,\ssSU}^{(1)}
+ 2 \bar{\chi}_{\!\ssP q_\mp,\ssSU}^{(1)}
  \bigg(
    \tw^2 \chi_{\!\ssP q_\mp,\ssY}^{(1)}
  - \sw^2 \chi_{\!q,\ssQED}^{(1)}
  \bigg)
\nn
\eq
\vspace{-.1cm}
\bq
[\bar{\chi}^2]_{_{\ssT u_-,\ssSU}}^{(1)}\!\! = - 4 L_{us}^2 - 4 L_{ts} L_{us},
\qquad\qquad
[\bar{\chi}^2]_{_{\ssT d_-,\ssSU}}^{(1)}\!\! = - 4 L_{ts}^2 - 4 L_{ts} L_{us},
\nn
\eq
\vspace{-.1cm}
\bq
[\bar{\chi}^2]_{_{\ssL u_-,\ssSU}}^{(1)} = - 4 L_{us}^2 - 4 L_{ts} L_{us},
\qquad
[\bar{\chi}^2]_{_{\ssL d_-,\ssSU}}^{(1)} = - 4 L_{ts}^2 - 4 L_{ts} L_{us},
\qquad
[\bar{\chi}^2]_{_{\ssL q_+,\ssSU}}^{(1)} = 0.
\eq

\vspace{.2cm}
\noindent
Finally, the pure two-loop quantities $\gamma_{\ssP q_\mp}^{(2)}$ are given by:
\bq
\gamma_{\ssP q_\pm}^{(2)}=
  \gamma_{\ssP q_\pm,\ssSU}^{(2)}
+ \tw^4\gamma_{\ssP q_\pm,\ssY}^{(2)}
- \sw^4\gamma_{q,\ssQED}^{(2)},
\qquad\qquad
\gamma_{u,\ssQED}^{(2)} =
{10400\over243},
\qquad\qquad
\gamma_{d,\ssQED}^{(2)} =
{8000\over243},
\nn
\eq
\bq
\gamma_{\ssT q_-,\ssSU}^{(2)} =
- {385\over9}
+ {11\over3}\pi^2,
\quad\;
\gamma_{\ssT q_-,\ssY}^{(2)} =
{52\over81},
\qquad\qquad
\gamma_{\ssL q_-,\ssSU}^{(2)} =
- {70\over3}
+ 2\pi^2,
\quad\;
\gamma_{\ssL q_-,\ssY}^{(2)} =
{520\over81},
\nn
\eq
\vspace{-.1cm}
\bq
\gamma_{\ssL q_+,\ssSU}^{(2)} =
- {35\over3}
+ \pi^2,
\qquad\qquad
\gamma_{\ssL u_+,\ssY}^{(2)} =
{1300\over81},
\qquad\qquad
\gamma_{\ssL d_+,\ssY}^{(2)} =
{676\over81}.
\eq

\vspace{.4cm}
\noindent
Throughout  this section the following notations have been used
\bqa
L &=& \ln\!\frac{\hat{s}}{M_W^2} - i\pi,
\qquad\!
L_{ts} = \ln\frac{-\hat{t}}{\hat{s}} + i\pi,
\qquad\!
L_{us} = \ln\frac{-\hat{u}}{\hat{s}} + i\pi,
\qquad\!
L_{ut} = \ln\frac{\hat{u}}{\hat{t}},
\qquad\!
L_{tu} = \ln\frac{\hat{t}}{\hat{u}},
\nn\\
L_{\!_\pm} &=&
\frac{-\hat{s}}{\hat{t}} \bigg( L_{us}^2 + \pi^2 \bigg)
\pm \frac{-\hat{s}}{\hat{u}} \bigg( L_{ts}^2 + \pi^2 \bigg),
\qquad\qquad
\Ltw = \ln\bigg(1-{M_W^2\over m_t^2}\bigg),
\qquad\qquad
L_\gamma = \ln\frac{\lambda^2}{M_Z^2},
\nn\\[+.2cm]
L_i\! &=& \ln\frac{M_i^2}{M_W^2},
\qquad
L_{\chi_{_i}} = \ln\frac{1\!-\!\beta_i}{1\!+\!\beta_i},
\qquad
\beta_i = -i\sqrt{4{M_W^2\over M_i^2}\!-\!1},
\qquad
w_i = \frac{M_i^2}{M_W^2},
\qquad
i= Z,H,t.
\qquad
\eqa
\section*{Appendix B}
In this appendix we present  the result for the one  and two-loop
corrections to the partonic cross section in NNLL approximation.
The differential cross sections are obtained from the amplitudes
given in the previous appendix through the relations:
\bqa
\frac{d\,\hat{\sigma}_{\!\ssT q_-}}{d\ct}\! &=&
\frac{N_c}{32\pi \hat{s}}\,
\overline{\sum_{{\scriptscriptstyle{\!S\!P\!I\!N\!}}}}\;\;
\overline{\sum_{\!\!\kappa=\pm1\!\!}}\;\;
\big|A_{\ssT q_-}\!\big|^2\!
=
\frac{\hat{s}\,\st^2}{256\pi}\bigg[
  \big|{\cal A}_{\ssT q_-}\!\big|^2 (1 \!+\! \ct^2)
- {\rm Re}\Big(\!{\cal A}_{\ssT q_-} {\cal B}_{\ssT q_-}^* \!\!\Big)
  \frac{\hat{s}}{2}\st^2\,\ct
+ \big|{\cal B}_{\ssT q_-}\!\big|^2 \frac{\hat{s}^2}{16}\,\st^4
\bigg],
\nn\\
\frac{d\,\hat{\sigma}_{\!\ssL q_\mp}}{d\ct} &=&
\frac{N_c}{32\pi \hat{s}}\,
\overline{\sum_{{\scriptscriptstyle{\!S\!P\!I\!N\!}}}}\;
\big|A_{\ssL q_\mp}\big|^2
=
\frac{\hat{s}\,\st^2}{128\pi} \big|{\cal A}_{\ssL q_\mp}\big|^2,
\qquad\qquad\qquad
\st= \sin\theta,
\quad
\ct= \cos\theta.
\eqa
The perturbative series~\eqn{sersig} for the cross section takes the form
\bq
{d\,\hat{\sigma}_{\ssP q_\mp} \over d\ct}=
  \left[
  1
+ \left( {\alpha \over 4\pi} \right) \dl_{\ssP q_\mp}^{(1)}
+ \left( {\alpha\over 4\pi} \right)^2 \dl_{\ssP q_\mp}^{(2)}
+ \ldots
\right] {d\,\hat{\sigma}_{\ssP q_\mp}^{(0)} \over d \ct}.
\qquad\qquad
\alpha= \frac{\al_e}{\sw^2},
\nn
\eq
\vspace{-.2cm}
\bq
{d\,\hat{\sigma}_{\ssT q_-}^{(0)} \over d \ct}\! =
N_c\frac{\alpha^2\pi}{16}\,\hat{s}\,\st^2\,(1 + \ct^2)\,
\big|{\cal A}_{\ssT q_-}^{(0)}\!\big|^2,
\qquad\qquad\qquad
{d\,\hat{\sigma}_{\ssL q_\mp}^{(0)} \over d \ct}\! =
N_c\frac{\alpha^2\pi}{8}\,\hat{s}\,\st^2\,
\big|{\cal A}_{\ssL q_\mp}^{(0)}\!\big|^2.
\eq
We  expand the corrections terms $\dl_{\ssP q_\mp}^{(n)}$ in powers of the large
logarithm ${\cal L}=\ln(\hat{s}/M_W^2)$
\bq
\delta_{\ssP q_\mp}^{(1)}=
  a_{\ssP q_\mp}^{(1)}\,{\cal L}^2
+ b_{\ssP q_\mp}^{(1)}\,{\cal L}
+ c_{\ssP q_\mp}^{(1)},
\qquad\qquad
\delta_{\ssP q_\mp}^{(2)}=
  a_{\ssP q_\mp}^{(2)}\,{\cal L}^4
+ b_{\ssP q_\mp}^{(2)}\,{\cal L}^3
+ c_{\ssP q_\mp}^{(2)}\,{\cal L}^2.
\eq
Numerically for the one-loop coefficients we get
\bqa
a_{\ssT u_-}^{(1)}
&=&
- 4.85\,,
\qquad\qquad\qquad
b_{\ssT u_-}^{(1)} =
  \bigg( - 6.77 + 4{\hat{u}\over\hat{t}} \bigg)l_u
+ 2.77\,l_t
+ 4.86\,,
\qquad\qquad\qquad\qquad\qquad
\nn\\
c_{\ssT u_-}^{(1)}
&=&
  \bigg(\! - 2.48 + 1.55 {\hat{u}\over\hat{t}} - 2.48 {\hat{u}^2\over\hat{t}^2}  \bigg)l_u^2
+ 3\,l_t^2
- 4 {s\over t}\,l_u\,l_t
+ \bigg( 5.25 - 4.97 {\hat{u}\over\hat{t}} \bigg)l_u
- 0.70\,l_t
- 3.24
\nn\\
&&
+ \,{(\hat{t}\!-\!\hat{u})\,\hat{u}\over \hat{u}^2\!+\!\hat{t}^2}\bigg[
    \bigg( 4.03 {\hat{u}\over\hat{t}} - 2.48 {\hat{u}^2\over\hat{t}^2}  \bigg)l_u^2
  - 3 {\hat{t}\over\hat{u}}\,l_t^2
  + \bigg( 4.55 - 4.97 {\hat{u}\over\hat{t}} \bigg)l_u
  - 2.48
  \bigg];
\nn\\[+.3cm]
a_{\ssT d_-}^{(1)}
&=&
- 5.00\,,
\qquad\qquad\qquad
b_{\ssT d_-}^{(1)}=
\bigg( - 7.38 + 4{\hat{t}\over\hat{u}} \bigg)l_t
+ 3.38\,l_u
+ 5.40\,,
\nn\\
c_{\ssT d_-}^{(1)}
&=&
  \bigg(\! - 2.48 + 1.55 {\hat{t}\over\hat{u}} - 2.48 {\hat{t}^2\over\hat{u}^2}  \bigg)l_t^2
+ 3\,l_u^2
- 4 {s\over u}\,l_u\,l_t
+ \bigg( 5.40 - 4.97 {\hat{t}\over\hat{u}} \bigg)l_t
- 0.85\,l_u
- 3.37
\nn\\
&&
+ \,{(\hat{u}\!-\!\hat{t})\,\hat{t}\over \hat{u}^2\!+\!\hat{t}^2}\bigg[
    \bigg( 4.03 {\hat{t}\over\hat{u}} - 2.48 {\hat{t}^2\over\hat{u}^2}  \bigg)l_t^2
  - 3 {\hat{u}\over\hat{t}}\,l_u^2
  + \bigg( 4.55 - 4.97 {\hat{t}\over\hat{u}} \bigg)l_t
  - 2.48
  \bigg];
\nn\\[+.3cm]
a_{\ssL u_-}^{(1)}
&=&
- 2.50\,,
\qquad\qquad\qquad\qquad
b_{\ssL u_-}^{(1)} =
- 4.97\,l_u
+ 0.97\,l_t
- 2.81\,,
\nn\\
c_{\ssL u_-}^{(1)}
&=&
  1.55{s\over t}\,l_u^2
- 0.55{s\over u}\,l_t^2
+ 0.24\,l_u
- 0.24\,l_t
+ 70.47\,;
\nn\\[+.3cm]
a_{\ssL d_-}^{(1)}
&=&
- 2.65\,,
\qquad\qquad\qquad\qquad
b_{\ssL d_-}^{(1)} =
- 5.18\,l_t
+ 1.18\,l_u
-2.27\,,
\nn\\
c_{\ssL d_-}^{(1)} &=&
  1.45{s\over u}\,l_t^2
- 0.45{s\over t}\,l_u^2
+ 0.30\,l_t
- 0.30\,l_u
+ 1.01\,;
\nn
\eqa
\vspace{-.9cm}
\bqa
a_{\ssL u_+}^{(1)}
&=&
- 1.25\,,
\qquad\qquad\qquad\qquad
b_{\ssL u_+}^{(1)} =
- 0.43\,l_t
+ 0.43\,l_u
- 6.69\,,
\qquad\qquad
\qquad\qquad\qquad\qquad
\nn\\
c_{\ssL u_+}^{(1)}
&=&
  0.20 {s\over t}\,l_u^2
- 0.20 {s\over u}\,l_t^2
- 0.11\,l_u
+ 0.11\,l_t
+ 46.23\,;
\nn\\[+.3cm]
a_{\ssL d_+}^{(1)}
&=&
- 1.20\,,
\qquad\qquad\qquad\qquad
b_{\ssL d_+}^{(1)} =
  0.22\,l_t
- 0.22\,l_u
- 6.85\,,
\nn\\
c_{\ssL d_+}^{(1)}
&=&
  0.10 {s\over u}\,l_t^2
- 0.10 {s\over t}\,l_u^2
- 0.05\,l_t
+ 0.05\,l_u
+ 46.31\,.
\eqa
\\[-.1cm]
The two-loop coefficients read
\\[-.4cm]
\bqa
a_{\ssT u_-}^{(2)} &=& 
11.76\,,
\qquad\qquad\qquad
b_{\ssT u_-}^{(2)} =
\bigg( 32.82 - 19.40 {\hat{u}\over\hat{t}} \bigg)l_u
- 13.42\,l_t
- 17.34\,,
\nn\\
c_{\ssT u_-}^{(2)}&=&
  \bigg(
  34.95 - 22.59 {\hat{u}\over\hat{t}} + 16.04 {\hat{u}^2\over\hat{t}^2} 
  \bigg)l_u^2
- 10.72\,l_t^2
+ \bigg(\! - 34.13 - 20.33 {\hat{u}\over\hat{t}} \bigg)l_u\,l_t
\nn\\
&&
+ \bigg(\! - 44.44 + 37.21 {\hat{u}\over\hat{t}} \bigg)l_u
+ 9.24\,l_t
- 28.46 + 39.48{\hat{u}^2\over\hat{t}^2}
\nn\\
&&
+ {(\hat{t}\!-\!\hat{u})\,\hat{u}\over \hat{u}^2\!+\!\hat{t}^2}\bigg[
    \bigg(\! 
    - 19.56 {\hat{u}\over\hat{t}} + 12.04 {\hat{u}^2\over\hat{t}^2}
    \bigg)l_u^2
  + 14.55 {\hat{t}\over\hat{u}}\,l_t^2
  + \bigg(\! - 22.06 + 24.08 {\hat{u}\over\hat{t}} \bigg)l_u
  + 12.04
  \bigg]\,;
\nn\\[+.3cm]
a_{\ssT d_-}^{(2)}
&=&
12.52\,,
\qquad\qquad\qquad
b_{\ssT d_-}^{(2)} =
\bigg( 36.94 - 20.01 {\hat{t}\over\hat{u}} \bigg)l_t
- 16.93\,l_u
- 20.89\,,
\nn\\
c_{\ssT d_-}^{(2)}&=&
  \bigg( 39.69 - 25.29 {\hat{t}\over\hat{u}} + 16.42 {\hat{t}^2\over\hat{u}^2}  \bigg)l_t^2
- 9.28\,l_u^2
+ \bigg(\! - 41.00 - 18.48 {\hat{t}\over\hat{u}} \bigg)l_u\,l_t
\nn\\
&&
+ \bigg(\! - 53.63 + 40.13 {\hat{t}\over\hat{u}} \bigg)l_t
+ 15.58\,l_u
- 23.56 + 39.48{\hat{t}^2\over\hat{u}^2}
\nn\\
&&
+ {(\hat{u}\!-\!\hat{t})\,\hat{t}\over \hat{u}^2\!+\!\hat{t}^2}\bigg[
    \bigg(\! - 20.18 {\hat{t}\over\hat{u}} + 12.42 {\hat{t}^2\over\hat{u}^2}  \bigg)l_t^2
  + 15.01 {\hat{u}\over\hat{t}}\,l_u^2
  + \bigg(\! - 22.77 + 24.85 {\hat{t}\over\hat{u}} \bigg)l_t
  + 12.42
  \bigg]\,;
\nn\\[+.3cm]
a_{\ssL u_-}^{(2)}
&=&
3.12\,,
\qquad\qquad\qquad\qquad
b_{\ssL u_-}^{(2)} =
  12.42\,l_u
- 2.42\,l_t
+ 10.54\,,
\qquad\qquad
\nn\\
c_{\ssL u_-}^{(2)} &=&
  \bigg(\! - 3.87{s\over t} + 12.34 \bigg)l_u^2
+ \bigg( 1.37{s\over u} + 0.47 \bigg)l_t^2
- 0.81\,l_u\,l_t
+ 23.93\,l_u
- 6.34\,l_t
- 274.80\,;
\nn\\[+.3cm]
a_{\ssL d_-}^{(2)}
&=&
3.52\,,
\qquad\qquad\qquad\qquad
b_{\ssL d_-}^{(2)} =
  13.76\,l_t
- 3.14\,l_u
+ 9.44\,,
\nn\\
c_{\ssL d_-}^{(2)}
&=&
  \bigg(\! - 3.85 {s\over u} + 13.44 \bigg)l_t^2
+ \bigg( 1.19 {s\over t} + 0.70 \bigg)l_u^2
- 2.14\,l_t\,l_u
+ 21.33\,l_t
- 5.91\,l_u
- 101.71\,;
\nn\\[+.3cm]
a_{\ssL u_+}^{(2)}
&=&
0.78\,,
\qquad\qquad\qquad\qquad
b_{\ssL u_+}^{(2)} =
  0.54\,l_t
- 0.54\,l_u
+ 10.11\,,
\nn\\
c_{\ssL u_+}^{(2)}&=&
  \bigg( 0.34 + 0.25 {\hat{u}\over\hat{t}} \bigg)l_u^2
+ \bigg(\! - 0.16 - 0.25 {\hat{t}\over\hat{u}} \bigg)l_t^2
- 0.19\,l_u\,l_t
- 2.30\,l_u
+ 2.30\,l_t
- 88.00\,;
\nn\\[+.3cm]
a_{\ssL d_+}^{(2)}
&=&
0.72\,,
\qquad\qquad\qquad\qquad
b_{\ssL d_+}^{(2)} =
- 0.26\,l_t
+ 0.26\,l_u
+ 10.03\,,
\nn\\
c_{\ssL d_+}^{(2)}
&=&
  \bigg(\! 0.14 + 0.12 {\hat{t}\over\hat{u}} \bigg)l_t^2
+ \bigg(\! - 0.10 - 0.12 {\hat{u}\over\hat{t}} \bigg)l_u^2
- 0.05\,l_u\,l_t
- 1.19\,l_t
+ 1.19\,l_u
- 85.17\,.
\eqa
\\[-.1cm]
Here $l_u= \ln(-\hat{u}/\hat{s})$ and
$l_t= \ln(-\hat{t}/\hat{s})$.

\section*{Appendix C}
In Ref.~\cite{Kuhn:2007ca} the contribution of the imaginary part of the
anomalous dimension matrix $\chi_T^{(1)}$ (given in Eq.~(\ref{trsu2}) above) 
has been missed in the numerical estimates. 
This contribution changes the NNLL two-loop correction
in the transverse boson production cross section. It  results in an additional
term 
\[
4\pi^2{x_-^2-x_+^2\over x_+^2}
\] 
in the coefficient of the quadratic logarithm in Eqs.~(14,~33)
of Ref.~\cite{Kuhn:2007ca}.


\begin{thebibliography}{99}


\bibitem{Fad}    V.S. Fadin, L.N. Lipatov, A.D. Martin, and M. Melles,
                 {Phys. Rev.} {D 61} (2000)  094002.

\bibitem{KPS}    J.H. K\"uhn, A.A. Penin, and V.A. Smirnov,
                 {Eur. Phys. J.}  {C 17} (2000) 97;
                 {Nucl. Phys. B (Proc. Suppl.)} {89} (2000) 94.

\bibitem{Mel1}   M. Melles,  {Phys. Rev.} {D 63} (2001) 034003; D 64
                 (2001) 014011.

\bibitem{DMP}    A. Denner, M. Melles, and  S. Pozzorini,
                 {Nucl. Phys.} {B 662} (2003) 299.

\bibitem{KMPS}   J.H. K\"uhn, S. Moch, A.A. Penin, and V.A. Smirnov,
                 {Nucl.Phys.} {B 616} (2001) 286, Erratum {\it  ibid.}
                 {B 648} (2003)  455.

\bibitem{FKPS}   B. Feucht, J.H. K\"uhn, A.A. Penin, and V.A. Smirnov,
                 Phys. Rev. Lett. {93} (2004) 101802.

\bibitem{JKPS}   B. Jantzen, J.H. K\"uhn, A.A. Penin, and V.A. Smirnov,
                 {Phys. Rev.} {D 72} (2005) 051301(R);
                 Nucl.\ Phys.\  B { 731} (2005) 188, [Erratum {\it
                   ibid.}   B { 752} (2006) 32.

\bibitem{Kuhn:1999de}
  J.~H.~K\"uhn and A.~A.~Penin,
  arXiv:hep-ph/9906545.

\bibitem{Ciafaloni:2000df}
  M.~Ciafaloni, P.~Ciafaloni and D.~Comelli,
  Phys.\ Rev.\ Lett.\  { 84} (2000) 4810.

\bibitem{Denner:2000jv}
  A.~Denner and S.~Pozzorini,
  Eur.\ Phys.\ J.\  C { 18} (2001) 461;
  A.~Denner and S.~Pozzorini,
  Eur.\ Phys.\ J.\  C { 21} (2001) 63;
  S.~Pozzorini,
  Nucl.\ Phys.\  B { 692} (2004) 135.



\bibitem{Man1}  J. Chiu, A. Fuhrer, R. Kelley, A.V. Manohar,
                Phys.\ Rev.\ D 77 (2008) 053004.

\bibitem{LemVel} M. Lemoine and M.J.G. Veltman,
                 Nucl.\ Phys.\  B {164}  (1980) 445.

\bibitem{Boh}    M.B\"ohm, A.Denner, T.Sack, W.Beenakker,
                 F.A. Berends, and H. Kuijf,
                 Nucl.\ Phys.\  B {304} (1988) 463.

\bibitem{FJZ}    J. Fleischer, F. Jegerlehner, and M. Zralek,
                 Z.\ Phys.\  C { 42} (1989) 409.

\bibitem{Bee}    W. Beenakker, A. Denner, S. Dittmaier, R. Mertig and T. Sack,
                 Nucl.\ Phys.\  B { 410}  (1993) 245.

\bibitem{Bee2}   W. Beenakker, F.A. Berends, A.P. Chapovsky, Nucl.\ Phys.\ B
{548} (1999) 3.


\bibitem{Jad}  S. Jadach et al., Comput. Phys. Commun. 140 (2001) 432;
               Phys. Rev. D 65 (2002) 093010;

\bibitem{Den2} A. Denner, S. Dittmaier, M. Roth and D. Wackeroth,
               Nucl. Phys. B 587 (2000) 67; Comput. Phys. Commun. 153 (2003)
462.

\bibitem{Den}    A. Denner, S. Dittmaier, M. Roth and L.H. Wieders,
                 Phys.\ Lett.\  B {612} (2005) 223; Nucl.\ Phys.\
                 B {724} (2005) 247.


\bibitem{BRV}   M. Beccaria, F.M. Renard, and C. Verzegnassi,
                Nucl.\ Phys.\  B { 663} (2003) 394.

\bibitem{Kuhn:2007ca}
  J.~H.~K\"uhn, F.~Metzler and A.~A.~Penin,
  Nucl.\ Phys.\  B { 795} (2008) 277
  [Erratum-ibid.\  { 818} (2009) 135].

\bibitem{ADP}    E. Accomando, A. Denner, and S. Pozzorini,  Phys. Rev. D 65 (2002) 073003.
 
\bibitem{ADK}    E. Accomando, A. Denner, and A. Kaiser, Nucl. Phys. B 706 (2005) 325. 

\bibitem{Man2}  J. Chiu, A. Fuhrer, R. Kelley, A.V. Manohar,
                Phys.\ Rev.\ D 80 (2009) 094013.

\bibitem{BHO} U. Baur, T. Han, and J. Ohnemus, Phys. Rev. D53 (1996) 1098. 

\bibitem{DKS} L.J. Dixon,  Z. Kunszt, and  A. Signer, Phys. Rev. D 60 (1999) 114037.

\bibitem{Gra}  M. Grazzini, JHEP 0601 (2006) 095. 

\bibitem{Binoth:2006mf}
  T.~Binoth, M.~Ciccolini, N.~Kauer {\it et al.},
  JHEP {\bf 0612 } (2006)  046.
  [hep-ph/0611170].

\bibitem{Sud}    V.V. Sudakov, {Zh. Eksp. Teor. Fiz.} {30} (1956) 87.

\bibitem{Jac}    R. Jackiw, {Ann. Phys.} {48} (1968) 292; {51} (1969) 575.

\bibitem{Mue}    A.H. Mueller {Phys. Rev.} {D 20} (19) 2037

\bibitem{Col}    J.C. Collins, {Phys. Rev.} {D 22} (1980) 1478;
                 Adv. Ser. Direct. High Energy Phys. 5 (1989) 573.

\bibitem{Sen1}   A. Sen, {Phys. Rev.} {D 24} (1981) 3281.

\bibitem{DJP}    A.~Denner, B.~Jantzen and S.~Pozzorini,
                 Nucl. Phys.  B { 761} (2007) 1.

\bibitem{FreTay} J. Frenkel and J.C. Taylor, {Nucl. Phys.}
                 {B 116} (1976) 185.

\bibitem{Sen2}   A. Sen, {Phys. Rev.} {D 28} (1983) 860.

\bibitem{Ste}    G. Sterman, {Nucl. Phys.} {B 281} (1987) 310.

\bibitem{BotSte} J. Botts and G. Sterman, {Nucl. Phys.} {B 325} (1989) 62.








\bibitem{Mel2}   M. Melles,   Phys. Rev.  D { 64} (2001) 014011.













\bibitem{Martin:2002dr}
  A.~D.~Martin, R.~G.~Roberts, W.~J.~Stirling and R.~S.~Thorne,
  Phys.\ Lett.\  B { 531} (2002) 216.

\bibitem{cuba}
  T.~Hahn,
  Comput.\ Phys.\ Commun.\  {168} (2005) 78.

\bibitem{BDJ}    
  M.B\"ohm, A.Denner and H. Joos,
  {\it ``Gauge theories of the strong and electroweak 
    interaction''}, Teubner, Stuttgart (2001).

\end{thebibliography}
\end{document}